\documentclass[11pt,a4paper]{article}
\usepackage{times}
\usepackage{a4wide}
\usepackage{amsfonts}
\usepackage{amssymb}
\usepackage{amsmath}
\usepackage{booktabs} % for much better looking tables
\usepackage{array}
\usepackage{ifpdf}
\ifpdf
\usepackage[pdftex,unicode,implicit]{hyperref}
\hypersetup{%
  pdftitle    = {Black holes and black strings of \texorpdfstring{$N=2$}{N=2}, 
\texorpdfstring{$d=5$}{d=5} supergravity in the H-FGK formalism},
  pdfkeywords = {supersymmetry, supergravity, very special geometry, black
    holes, BPS solutions},
  pdfauthor   = {Patrick Meessen, Tomas Ort\'{\i}n, Jan Perz, C.S. Shahbazi},
  plainpages  = true,
  colorlinks  = true,
  citecolor   = blue,
  urlcolor    = red,
  linkcolor   = black
}
\newcommand{\hepth}[1]{{\tt
\href{http://www.arXiv.org/abs/hep-th/#1}{hep-th/#1}}}

\newcommand{\arxiv}[1]{{\tt
\href{http://www.arXiv.org/abs/#1}{arXiv:#1}}}
\else
  \usepackage[dvips]{graphicx}
  \usepackage[unicode,implicit]{hyperref}
  \newcommand{\hepth}[1]{{\tt hep-th/#1}}

  \newcommand{\arxiv}[1]{{\tt arXiv:#1}}
\fi
\usepackage{tikz}
\newcommand{\FPAUO}[2]{
\tikz[scale=.13,
         Uniovi/.style={color=gray, fill=gray}
 ] {
 \fill[Uniovi] (0,0) circle (10);
 \fill[white] (0,7) circle (1.5);
 \draw[Uniovi] (-2,7.5) rectangle (2,5.5);
 \fill[white] (-0.3,6.6) rectangle (0.3,0);   % 1.7 cm 
 \fill[white] ( -0.9,6.2) rectangle (.9 ,5.6);
 \fill[white] (-1.4, 5.2) rectangle (1.4, 4.6);
 \fill[white] (0,0) ellipse (3.5 and 4);
 \fill[Uniovi] (-2.5,0.3) rectangle (2.5,-0.3);
 \fill[Uniovi] (-2,2.3) rectangle (2,1.7);
 \fill[Uniovi] (-2,-2.3) rectangle (2,-1.7);
 \fill[white] (-4.5,5.5) rectangle (-2.7,4.9);
 \fill[white] (-3.9,6.1) rectangle (-3.3,4.3);
 \fill[white] (4.5,5.5) rectangle (2.7,4.9);
 \fill[white] (3.9,6.1) rectangle (3.3,4.3);
 \foreach \x in { 0,..., 3 }
   \foreach \y in { 0,...,\x}
    {
     \fill[white] (-6-\x*0.7+\y*1.4,3.5-\x *1.97) -- (-5.6-\x*0.7+\y*1.4,2.4-\x *1.97) -- (-6.4-\x*0.7+\y*1.4,2.4-\x *1.97) -- cycle;
     \fill[white] (6-\x*0.7+\y*1.4,3.5-\x *1.97) -- (5.6-\x*0.7+\y*1.4,2.4-\x *1.97) -- (6.4-\x*0.7+\y*1.4,2.4-\x *1.97) -- cycle;
   };
 \draw (0,-6) node[
                               text centered, 
                               color=white, 
                               font={\fontsize{8}{4}\sffamily\selectfont}
                             ] {FPAUO-#1/#2};
}} 
\makeatletter
\@addtoreset{equation}{section}
\makeatother

\begin{document}

~\vspace{-2cm}
\begin{flushright}
\small
\FPAUO{12}{01}\\
IFT-UAM/CSIC-11-102\\
\texttt{arXiv:1204.0507 [hep-th]}\\
April 2\textsuperscript{nd}, 2012\\
\normalsize
\end{flushright}

\vspace{2cm}

\begin{center}

  {\Large {\bf Black holes and black strings of $N=2$, $d=5$ supergravity\\in
      the H-FGK formalism}\par}

\vspace{2cm}

\renewcommand{\thefootnote}{\alph{footnote}}
{\sl\large Patrick Meessen$^{\dagger}$}%
\footnote{{\tt meessenpatrick [at] uniovi.es}},
{\sl\large Tom\'{a}s Ort\'{\i}n$^{\diamond}$}%
\footnote{{\tt Tomas.Ortin [at] csic.es}},
{\sl\large Jan Perz$^{\diamond}$}%
\footnote{{\tt Jan.Perz [at] uam.es}}
{\sl\large and C.~S.~Shahbazi$^{\diamond}$}%
\footnote{{\tt Carlos.Shabazi [at] uam.es}}
\renewcommand{\thefootnote}{\arabic{footnote}}

\vspace{.5cm}

${}^{\dagger}${\it HEP Theory Group, Departamento de F\'{\i}sica, Universidad de Oviedo\\
  Avda.~Calvo Sotelo s/n, 33007 Oviedo, Spain}\\

\vspace{.4cm}

${}^{\diamond}${\it Instituto de F\'{\i}sica Te\'orica UAM/CSIC\\
C/ Nicol\'as Cabrera, 13--15, 28049 Madrid, Spain}\\

\vspace{2cm}

%%%%%%%%%%%%%%%%%%%%%%%%%%%%%%%%%%%%%%%%%%%%%%%%%%%%%%%%%%%%%%%%%%%%%%

{\bf Abstract}

\end{center}

\begin{quotation}
  \small We study general classes and properties of extremal and non-extremal
  static black-hole solutions of $N=2$, $d=5$ supergravity coupled to vector
  multiplets using the recently proposed H-FGK formalism, which we also extend
  to static black strings. We explain how to determine in general the
  integration constants and physical parameters of the black-hole and
  black-string solutions. We derive some model-independent statements,
  including the transformation of non-extremal flow equations to the form of
  those for the extremal flow. We apply our methods to the construction of
  example solutions (among others a new extremal string solution of heterotic
  string theory on $K_3 \times S^1$) and analyze their properties. In all the
  cases studied the product of areas of the inner and outer horizon of a
  non-extremal solution coincides with the square of the moduli-independent
  area of the horizon of the extremal solution with the same charges.
\end{quotation}

\newpage
%%%%%%%%%%%%%%%%%%%%%%%%%%%%%%%%%%%%%%%%%%%%%%%%%%%%%%%%%%%%%%%%%%%%%%
%%%%%%%%%%%%%%%%%%%%%%%%%%%%%%%%%%%%%%%%%%%%%%%%%%%%%%%%%%%%%%%%%%%%%%
%%%%%%%%%%%%%%%%%%%%%%%%%%%%%%%%%%%%%%%%%%%%%%%%%%%%%%%%%%%%%%%%%%%%%%
%%%%%%%%%%%%%%%%%%%%%%%%%%%%%%%%%%%%%%%%%%%%%%%%%%%%%%%%%%%%%%%%%%%%%%
\pagestyle{plain}
%%%%%%%%%%%%%%%%%%%%%%%%%%%%%%%%%%%%%%%%%%%%%%%%%%%%%%%%%%%%%%%%%%%%%%
%%%%%%%%%%%%%%%%%%%%%%%%%%%%%%%%%%%%%%%%%%%%%%%%%%%%%%%%%%%%%%%%%%%%%%
%%%%%%%%%%%%%%%%%%%%%%%%%%%%%%%%%%%%%%%%%%%%%%%%%%%%%%%%%%%%%%%%%%%%%%
%%%%%%%%%%%%%%%%%%%%%%%%%%%%%%%%%%%%%%%%%%%%%%%%%%%%%%%%%%%%%%%%%%%%%%
%%%%%%%%%%%%%%%%%%%%%%%%%%%%%%%%%%%%%%%%%%%%%%%%%%%%%%%%%%%%%%%%%%%%%%

\tableofcontents

%%%%%%%%%%%%%%%%%%%%%%%%%%%%%%%%%%%%%%%%%%%%%%%%%%%%%%%%%%%%%%%%%%%%%%
%%%%%%%%%%%%%%%%%%%%%%%%%%%%%%%%%%%%%%%%%%%%%%%%%%%%%%%%%%%%%%%%%%%%%%
%%%%%%%%%%%%%%%%%%%%%%%%%%%%%%%%%%%%%%%%%%%%%%%%%%%%%%%%%%%%%%%%%%%%%%
%%%%%%%%%%%%%%%%%%%%%%%%%%%%%%%%%%%%%%%%%%%%%%%%%%%%%%%%%%%%%%%%%%%%%%

\section{Introduction}

Static, spherically symmetric black-hole solutions of $N=2$ supergravity can be
conveniently studied in the effective black-hole potential formalism originally
developed by Ferrara, Gibbons and Kallosh \cite{Ferrara:1997tw} in four
dimensions, later extended to arbitrary dimensions \cite{Meessen:2011bd}, as
well as to $p$-branes \cite{Martin:2012bi}. This is especially true for
supersymmetric extremal solutions, where expressing the effective potential by
the central charge leads to the derivation of first-order flow equations for
the scalars (implied by the Killing spinor equations), whose attractor fixed
points (corresponding to the sets of values of scalars on the event horizon)
\cite{Ferrara:1995ih} are determined by critical points of the central charge.

Even though not all extremal black holes are supersymmetric
\cite{Khuri:1995xq,Ortin:1996bz} and all non-extremal black holes break all
supersymmetries, it turns out \cite{Miller:2006ay} that analogous flow
equations may be derived for four- \cite{Ceresole:2007wx} and five-dimensional
extremal black holes \cite{Cardoso:2007ky}, as well as for non-extremal
$p$-branes \cite{Janssen:2007rc} and black holes \cite{Perz:2008kh}. This fact
alone already hints at a possibility that perhaps non-supersymmetric solutions
could be obtained in a similar way to the supersymmetric ones. Indeed, at least
in a class of four- \cite{Galli:2011fq} and five-dimensional
\cite{Mohaupt:2010fk,Meessen:2011bd} black holes and $p$-branes
\cite{Martin:2012bi}, the known supersymmetric solution can be deformed to a
unique non-extremal solution, from which both supersymmetric and
non-supersymmetric extremal solutions are recovered in the different limits in
which the non-extremality parameter vanishes. This is, as far as we know, the
only systematic method for constructing general extremal non-supersymmetric
black-hole solutions, in particular when the black-hole potential has flat
directions and the values of the scalar fields on the horizon have some
dependence on the asymptotic values \cite{Galli:2011fq}.

That a deformation from a supersymmetric to a non-extremal solution must be
possible and that all static solutions with spherical symmetry can be treated
in the same manner, becomes clear in a new set of $H$-variables introduced in
the 5-dimensional case in \cite{Mohaupt:2009iq,Mohaupt:2010fk}\footnote{A
  different derivation specific for $N=2$ $d=5$ supergravity theories was also
  given in \cite{Meessen:2011aa}.} and in the 4-dimensional case in
\cite{Mohaupt:2011aa,Meessen:2011aa}\footnote{Again, the derivation of
  \cite{Meessen:2011aa} makes heavy use of the formalism of $N=2$, $D=4$
  supergravity.} in which all static, spherically symmetric black-hole
solutions of a given model take the same, universal functional form,
irrespective of supersymmetry or extremality, although the radial profile of
the $H$-variables themselves will be different for the different kinds of
solutions. These variables arise naturally in the classification of the
timelike supersymmetric solutions of these theories
\cite{Gauntlett:2002nw,Gauntlett:2004qy,Meessen:2006tu}, to which the
supersymmetric black holes belong, but also occur in the classification of the
timelike supersymmetric solutions of more general theories (with
hypermultiplets \cite{Huebscher:2006mr,Bellorin:2006yr}, gaugings
\cite{{Huebscher:2007hj}} or both \cite{kn:superzin,Bellorin:2007yp} or with
both and additionally tensor multiplets \cite{Bellorin:2008we}) and transform
linearly under the duality transformations (subgroups of $Sp(2n+2,\mathbb{R})$
in $d=4$ and $SO(n+1)$ in $d=5$, for $n$ vector multiplets). These variables
replace the scalars and the metric function of the theory that appear on
different footing in the effective action and their use should, in principle,
simplify and systematize the task of constructing explicit solutions and
general results.

In the present work we use the 5-dimensional version of this formalism to ask
general questions about the black-hole solutions of $N=2$, $d=5$ supergravity
and to construct some families of solutions. Furthermore, profiting from the
recent extension of the FGK formalism to $p$-branes in any dimension, we extend
the H-FGK formalism to cover the case of black strings in these theories,
introducing new $H$-variables inspired by the classification of the null
supersymmetric solutions of $N=2$, $d=5$ supergravities
\cite{Chamseddine:1999qs,Gauntlett:2002nw,Gauntlett:2004qy,Bellorin:2006yr,Bellorin:2008we}.
We then study the resulting system as we do with the one for black holes.

We start by reviewing the H-FGK formalism for black holes of $N=2$, $d=5$
supergravity coupled to vector multiplets in
section~\ref{sec-H-var}. Following \cite{Meessen:2011aa}, we introduce the
basic definitions concerning the theories we deal with, the metric ansatz and
the $H$-variables. We show how we can get the metric that covers the region
lying between the inner (Cauchy) horizon and the singularity (not discussed in
\cite{Meessen:2011bd,Martin:2012bi}) from the one that covers the exterior of
the outer (event) horizon in the present 5-dimensional case. This will allow
us to compute the ``entropy'' and ``temperature'' associated with the inner
horizon\footnote{The inner horizon is also reached, albeit in a different way,
  in ref.~\cite{Mohaupt:2010fk}.}.

In section~\ref{sec-d5extremal} we will apply the formalism just discussed to
the study of extremal (BPS and non-BPS) black holes under the reasonable
assumption that, for extremal black holes, all the $H$-variables are harmonic
functions in the transverse space, defined by two integration constants. We
recover the results obtained in \cite{Mohaupt:2010fk} and find some new
ones. We study how these integration constants can be determined as functions
of the physical parameters in general, finding that half of them are always
determined by the asymptotic values of the scalars, that can be fixed at
will. The other half play the r\^{o}le of ``fake charges'' and many physical
quantities (mass, entropy) are determined by the \textit{fake central charge}
(or superpotential) constructed by the standard formula with the charges
replaced by the fake charges. In the extremal case the fake charges can be
determined by extremization of the black-hole potential on the horizon, like in
the original FGK formulation, but with the actual black-hole potential now
understood as a function of the fake and physical charges, rather than of
scalars and physical charges. The first-order flow equations for extremal black
holes are constructed in section~\ref{sec:FirstOrder} using the simple
procedure proposed in \cite{Ortin:2011vm}, which is valid for
non-supersymmetric cases as well. The equations of motion of an extremal black
hole are reproduced when the \textit{fake black-hole potential} (a function of
scalars and fake charges) is equal to the true one.

We then go on to study the non-extremal case in section~\ref{sec-d5ansatz},
adopting for the $H$-variables the exponential or hyperbolic ansatz of
\cite{Mohaupt:2010fk,Galli:2011fq,Meessen:2011bd}. We show how the relation
between extremization of the black-hole potential and attractor behavior for
the scalars and the relation between entropy and black-hole potential on the
horizon are modified in the non-extremal case. In section \ref{sec:1stNonExt}
we demonstrate how the first-order equations for non-extremal black holes can
be brought to the form of the extremal flow.

Explicit solutions are given in section~\ref{sec-d5examples}. The examples that
we analyze include the general non-extremal black holes with constant scalars,
in section~\ref{sec-d5nonextremaldoublyextremal} (found earlier in
\cite{Mohaupt:2010fk} in a different way), the $STU$ model, in
section~\ref{sec-STU}, which we solve paying particular attention to the
possible signs of the charges, and in section~\ref{sec-Jordan} the models of
the reducible Jordan sequence, whose black-hole potential has flat directions
and whose values of scalars on the horizon, in some non-supersymmetric cases,
are not completely fixed by the charges.

Finally, in section~\ref{sec:StringFGK} we generalize this approach to black
strings, using the extension, recently constructed in \cite{Martin:2012bi}, of
the FGK formalism to $p$-branes, and introducing dual $H$-variables (which we
shall denote by $K$). Our study of this case follows what we did for the black
holes in the previous sections: we find the general solutions for non-extremal
black strings with constant scalars for any $N=2$, $d=5$ supergravity theory in
section~\ref{sec:DoubleBlackStrings}, derive flow equations for black strings
in section~\ref{sec:FlowString}, and construct explicitly the extremal black
strings of the pure and the heterotic $STU$ model in
section~\ref{sec-extremalstrings}.

Section~\ref{conclusions} contains our conclusions.

%%%%%%%%%%%%%%%%%%%%%%%%%%%%%%%%%%%%%%%%%%%%%%%%%%%%%%%%%%%%%%%%%%%%%%
%%%%%%%%%%%%%%%%%%%%%%%%%%%%%%%%%%%%%%%%%%%%%%%%%%%%%%%%%%%%%%%%%%%%%%
%%%%%%%%%%%%%%%%%%%%%%%%%%%%%%%%%%%%%%%%%%%%%%%%%%%%%%%%%%%%%%%%%%%%%%
%%%%%%%%%%%%%%%%%%%%%%%%%%%%%%%%%%%%%%%%%%%%%%%%%%%%%%%%%%%%%%%%%%%%%%
%%%%%%%%%%%%%%%%%%%%%%%%%%%%%%%%%%%%%%%%%%%%%%%%%%%%%%%%%%%%%%%%%%%%%%

\section{The H-FGK formalism in five dimensions}
\label{sec-d5}
%%%%%%%%%%%
\subsection{H-variables}
\label{sec-H-var}
%%%%%%%%%%%
We start by recalling the salient points of ref.~\cite{Meessen:2011aa}. $N=2$,
$d=5$ supergravity \cite{Gunaydin:1983rk} coupled to $n$ vector multiplets
contains, apart from the metric, $n$ scalar fields $\phi^{x}$ ($x=1,\ldots ,n$)
and $n+1$ vector fields $A^{I}$ ($I=0,\ldots n$). The coupling between these
fields is specified by real special geometry, which in itself can be formulated
in terms of a constant completely symmetric real tensor $C_{IJK}$ and a section
$h^{I}(\phi )$ that obeys the fundamental constraint
\begin{equation}
  \label{eq:34}
  \mathcal{V}(h) = C_{IJK} h^{I} h^{J} h^{K} = 1\, .
\end{equation}
If we then define the derived objects
\begin{equation}
  \label{eq:35}
  h_{I} \equiv C_{IJK}h^{J}h^{K}\, ,\qquad
  h^{I}_{x} \equiv -\sqrt{3}\ \frac{\partial h^{I}}{\partial \phi^{x}} \qquad\mbox{and}\qquad
  h_{Ix} \equiv \sqrt{3}\ \frac{\partial h_{I}}{\partial \phi^{x}} \, , 
\end{equation}
we can see that they satisfy the following relations
\begin{equation}
  \label{eq:30}
  h^{I}h_{I} = 1 \qquad\mbox{and}\qquad
  h^{I}h_{Ix} = h_{I}h^{I}_{x} = 0 \, .
\end{equation}
The metric on the scalar manifold, $\mathit{g}_{xy}$, and the vector kinetic
matrix, $\mathit{a}_{IJ}$, are given by
\begin{equation}
  \label{eq:36}
  \mathit{g}_{xy} = h_{Ix}h^{I}_{y} \qquad\mbox{and}\qquad
  \mathit{a}_{IJ} = 3h_{I}h_{J} - 2C_{IJK}h^{K} = h_{I}h_{J} + h_{Ix}h_{J}^{x}\, .
\end{equation}
With these definitions we can write the bosonic part of the action for $N=2$,
$d=5$ supergravity coupled to $n$ vector supermultiplets as
\begin{equation}
  \label{eq:37}
   \mathcal{I}_{5} =  \int_{5}\left(
      R \star\! 1 + \tfrac{1}{2}\mathit{g}_{xy}\,d\phi^{x}\wedge\star d\phi^{y}
       - \tfrac{1}{2} \mathit{a}_{IJ}F^{I}\wedge\star F^{J}
      + \tfrac{1}{3\sqrt{3}}C_{IJK}F^{I}\wedge F^{J}\wedge A^{K}
  \right) .
\end{equation}
Having briefly detailed the relevant physical theory that we want to consider,
we can discuss the FGK formalism.
%%%%%%%%%%%%%%%
\par
The starting point of the FGK formalism in 5 dimensions is the ansatz for a
spherically symmetric metric describing the exterior of the event horizon of a
generic 5-dimensional black hole, namely
\begin{equation}
\label{eq:32}
ds^{2} 
= 
e^{2U(\rho)}dt^{2} - e^{-U(\rho)}\left(
\frac{\mathcal{B}^{3}}{4\sinh^{3}(\mathcal{B}\rho)}\, d\rho^{2}  
+\frac{\mathcal{B}}{\sinh (\mathcal{B}\rho)}\, d\Omega^{2}_{(3)}
       \right) ,
\end{equation}
where $d\Omega^{2}_{(3)}$ is the round metric on the 3-sphere of unit radius
and $\mathcal{B}$ is the so-called {\em non-extremality parameter}, meaning
that extremal solutions are obtained as the $\mathcal{B}\rightarrow 0$ limit.
In the employed coordinate system the asymptotic region lies at $\rho = 0$,
whereas the putative horizon is located at $\rho \rightarrow\infty$: in order
for the metric (\ref{eq:32}) to describe a non-extremal black hole, the
function $U$ must have the following limiting behavior
\begin{equation}
  \label{eq:33}
  \lim_{\rho\rightarrow\infty} e^{-U} = e^{\mathcal{B}\rho} \, , 
\end{equation}
which ensures that the limiting spacetime is a 2-dimensional Rindler space
times a 3-sphere.
\par
Although this was not realized in \cite{Meessen:2011bd}, the same general
metric describes the interior of the inner (Cauchy) horizon as well, just as it
happens in $d=4$ dimensions \cite{Galli:2011fq}, although in this case it is
more difficult to see. Given a regular solution of the above form describing
the exterior of the black hole for $\rho\in (0,+\infty)$, we can obtain the
metric that describes the interior of the inner horizon by
transforming\footnote{This is \textit{not} a coordinate transformation,
  because, among other reasons, it relates the metric in two different,
  disjoint patches of the spacetime.} that metric according to
\begin{equation}
\rho \longrightarrow -\varrho\, ,
\hspace{1cm}
e^{-U(\rho)} \longrightarrow -e^{-U(-\varrho)}\, .
\end{equation}
The new metric has the same general form in terms of the coordinate which now
takes values in the range $\varrho\in (\varrho_{\rm sing},+\infty)$ because the
metric will generically hit a singularity before $\varrho$ reaches $0$: if the
original $e^{-U}$ is always finite for positive values of $\rho$, the
transformed one will have a zero for some finite positive value of $\varrho$,
as we will see in the examples.
\par
Being interested in spherically symmetric solutions black hole solutions we
take $\phi^{x}=\phi^{x}(\rho )$ and can solve the vector field equations of
motion by putting
\begin{equation}
  \label{eq:38}
  F^{I} = -\sqrt{3}\,e^{2U}\mathit{a}^{IJ}\,q_{J}\,dt\wedge d\rho \, ,
\end{equation}
where the $q$'s are the electric charges.  Using the ans\"atze (\ref{eq:32})
and (\ref{eq:38}) in the remaining equations of motion, we see that they all
reduce to the following equations
\begin{align}
  \label{eq:39}
  \ddot{U}  + e^{2U} V_{\rm bh}(\phi ,q) & = 0\, , \\
  \label{eq:39b}
  \ddot{\phi}^{x}  + \Gamma_{yz}{}^{x}\dot{\phi}^{y}\dot{\phi}^{z}  + 
     \tfrac{3}{2}\ e^{2U} \partial^{x}V_{\rm bh}(\phi ,q) & = 0\, ,\\
  \label{eq:39c}
  \dot{U}^{2} + \tfrac{1}{3}\mathit{g}_{xy}\dot{\phi}^{x}\dot{\phi}^{y}
          + e^{2U}V_{\rm bh}(\phi ,q)  - \mathcal{B}^{2} & = 0\, ,
\end{align}
where we used the over-dot to denote derivation with respect to $\rho$ and we
defined the {\em black hole potential} by
\begin{equation}
\label{eq:40}
V_{\rm bh}(\phi ,q)  \equiv -\mathit{a}^{IJ}q_{I}q_{J} 
= 
-\mathcal{Z}_{\rm e}^{2}  - 3\,\partial_{x}\mathcal{Z}_{\rm e}\,\partial^{x}\mathcal{Z}_{\rm e} \, ,
\end{equation}
and in the last step defined the {\em (electric) central charge} by
$\mathcal{Z}_{\rm e}=\mathcal{Z}_{\rm e}(\phi ,q)\equiv h^{I}q_{I}$.
\par
The equations (\ref{eq:39}) and (\ref{eq:39b}) can be obtained from the FGK
effective action
\begin{equation}
\label{eq:FGKd5}
\mathcal{I}[U,\phi^{x}] = \int d\rho 
\left(
   \dot{U}^{2}
    + \mathit{a}^{IJ}\dot{h}_{I}\dot{h}_{J}
    - e^{2U}V_{\rm bh}(\phi ,q)
    + \mathcal{B}^{2}
\right) ,
\end{equation}
where we made use of eqs.~(\ref{eq:35}) and (\ref{eq:36}) to cast it into a
more suitable form.  Given this action, eq.~(\ref{eq:39c}) can be interpreted
as the constraint that the Hamiltonian corresponding to eq.~(\ref{eq:FGKd5}) be
zero.
\par
At this point we introduce two new sets of variables, $\tilde{H}^{I}$ and
$H_{I}$, related to the original ones ($U,\phi^{x}$) by
\begin{align}
e^{-U/2}h^{I}(\phi) & \equiv  \tilde{H}^{I}\, ,\\
e^{-U}h_{I}(\phi) & \equiv  H_{I}\, ,  
\end{align}
\noindent
and two new functions $\mathsf{V}$ and $\mathsf{W}$ 
\begin{equation}
\mathsf{V}(\tilde{H}) \equiv  C_{IJK}\tilde{H}^{I}\tilde{H}^{J}\tilde{H}^{K}
\, ,
\qquad
\mathsf{W}(\tilde{H}) = 2\,\mathsf{V}(\tilde{H})\, ,
\end{equation}
\noindent
but which are not constrained. Using the homogeneity properties of these
functions we find that
\begin{align}
e^{-\frac{3}{2}U} 
& =
\tfrac{1}{2} \mathsf{W}(H)\, ,
\\
h_{I} 
& =
(\mathsf{W}/2)^{-2/3}H_{I}\, , 
\\
h^{I} 
& =
(\mathsf{W}/2)^{-1/3}\tilde{H}^{I}\, . 
\end{align}

We can use these formulae to perform the change of variables in the effective
action for static, spherically symmetric black holes of $N=2,d=5$ supergravity
\cite{Meessen:2011bd}, which can be rewritten in the convenient form
\begin{equation}
\mathcal{I}[U,\phi^{x}] 
= \int d\rho 
\left(
\dot{U}^{2}
+a^{IJ}\dot{h}_{I}\dot{h}_{J}
+e^{2U}a^{IJ}q_{I}q_{J}
+\mathcal{B}^{2}
\right) .   
\end{equation}
\noindent
Thanks to the identity
\begin{equation}
a^{IJ} = -\frac{3}{2}\left(\frac{\mathsf{W}}{2}\right)^{4/3} 
\partial^{I}\partial^{J}\log\mathsf{W}
\end{equation}
\noindent
the above action, in terms of the $H_{I}$ variables, becomes
\begin{equation}
\label{eq:effectived52}
-\tfrac{3}{2}\mathcal{I}[H] 
= \int d\rho 
\left(
\partial^{I}\partial^{J}
\log\mathsf{W}\, 
\bigl( \dot{H}_{I}\dot{H}_{J}+q_{I}q_{J}\bigr)
-\tfrac{3}{2}\mathcal{B}^{2}
\right) .   
\end{equation}

The equations of motion derived from the effective action are
\begin{equation}
\label{eq:eomsd5}
\partial^{K}\partial^{I}\partial^{J}
\log\mathsf{W}\left(
      H_{I}\ddot{H}_{J}
    - \dot{H}_{I}\dot{H}_{J}
    + q_{I}q_{J}
\right) = 0\, .
\end{equation}
\noindent
Multiplying these equations by $\dot{H}_{K}$ we get $\dot{\mathcal{H}}=0$, the
Hamiltonian constraint
\begin{equation}
\label{eq:hamiltoniand5}
\mathcal{H} \equiv   
\partial^{I}\partial^{J}
\log\mathsf{W}\, 
\bigl( \dot{H}_{I}\dot{H}_{J}-q_{I}q_{J}\bigr)
+\tfrac{3}{2}\mathcal{B}^{2}
=0\, ,
\end{equation}
\noindent
where the integration constant has been set to
$\tfrac{3}{2}\mathcal{B}^{2}$. Multiplying the equations of motion by $H_{K}$
we obtain
\begin{equation}
\label{eq:d5Ueq}
\partial^{I}\log\mathsf{W}\,\ddot{H}_{I}= \tfrac{3}{2}\mathcal{B}^{2}\, ,  
\end{equation}
\noindent
which is the equation of $U$ expressed in the new variables.

In the next subsections we shall use this formalism to study general families
of solutions.

%%%%%%%%%%%%%%%%%%%%%%%%%%%%%%%%%%%%%%%%%%%%%%%%%%%%%%%%%%%%%%%%%%%%%%
%%%%%%%%%%%%%%%%%%%%%%%%%%%%%%%%%%%%%%%%%%%%%%%%%%%%%%%%%%%%%%%%%%%%%%

\subsection{Extremal black holes}
\label{sec-d5extremal}

In the extremal black-hole case, $\mathcal{B}=0$, we expect the $H_{I}$ to be
harmonic functions on the transverse $\mathbb{R}^{4}$ space, {\em
  i.e.\/}~linear functions of $\rho$
\begin{equation}
\label{eq:harmonic}
  H_{I}=A_{I} +B_{I}\rho\, ,  
\end{equation}
\noindent
where the integration constants $A_{I}$ and $B_{I}$ are functions of the
physical parameters (electric charges $q_{I}$ and asymptotic values of the
scalars $\phi^{x}_{\infty}$) to be determined by requiring that the equations
of motion are satisfied everywhere, {\em i.e.\/}
\begin{align}
\label{eq:eomextremeansatz}
\partial^{K}\partial^{I}\partial^{J}
\log\mathsf{W}(H) 
\left( B_{I}B_{J}-q_{I}q_{J}\right)
& = 0\, ,\\
\label{eq:hamilextremeansatz}
\partial^{I}\partial^{J}
\log\mathsf{W}(H) 
\left( B_{I}B_{J}-q_{I}q_{J}\right)
& = 0\, ,
\end{align}
\noindent
and the physical fields are correctly normalized at spatial infinity
($\rho\rightarrow 0$). We study these conditions first.

Asymptotic flatness requires that $\mathsf{W}(H(0))=2$, {\em
  i.e.\/}\footnote{$\mathsf{W}(A)$ and other analogous expressions are to be
  understood as the functions one obtains when replacing the $H_{I}$ by the
  constants $A_{I}$.}
\begin{equation}
\label{eq:asympflatd5}
\mathsf{W}(A)=2\, .  
\end{equation}
\noindent
In order to study the asymptotic behavior of the scalars it is convenient to
introduce the following model-independent definition for the $n$ physical
scalars of a generic $N=2,d=5$ theory:\footnote{We use the symbol $\varphi$ to
  distinguish scalars defined by eq.~(\ref{eq:scalardef}) from physical scalars
  in an arbitrary, possibly different parametrization, which we denote by
  $\phi$. For an explicit example see the paragraph after
  eq.~(\ref{eq:hvarphi}).}
\begin{equation}
\label{eq:scalardef}
\varphi_{x} \equiv \frac{h_{x}}{h_{0}}\, .
\end{equation}
\noindent
The possible values of these scalars have to be determined model by model. We
will ignore in this general discussion all issues related to their possible
signs, singular values etc. Other choices amount to field redefinitions but
eq.~(\ref{eq:scalardef}) allows us to write the solutions for the physical
scalars in terms of the functions $H_{I}$ in a generic way as
\begin{equation}
\varphi_{x} = \frac{H_{x}}{H_{0}}\, .
\end{equation}
\noindent
Hence, the asymptotic values of all $H_{I}$, $I\ne 0$, are given by 
\begin{equation}
\label{eq:Avarphiinfty}
A_{x} = \varphi_{x\, \infty\,}A_{0}\, .  
\end{equation}

Now, defining for convenience $\varphi_{0}\equiv 1$, we can write
\begin{equation}
\mathsf{W}(\varphi) = \mathsf{W}(H/H_{0})= H_{0}^{-3/2} \mathsf{W}(H)\, ,  
\end{equation}
\noindent
and taking into account the normalization at spatial infinity, we find that
\begin{equation}
\label{eq:AIvarphi}
A_{0} = [\mathsf{W}(\varphi_{\infty})/2]^{-2/3}\, .  
\end{equation}
\noindent
Summarizing, we have shown that, in any model, the constants $A_{I}$ are given
in terms of the asymptotic values of the scalars by 
\begin{align}
\label{eq:asympconstantsd5-1}
A_{I}  & = \varphi_{I\, \infty}[\mathsf{W}(\varphi_{\infty})/2]^{-2/3}\, ,
\\
\label{eq:asympconstantsd5-2}
\varphi_{I\, \infty} & = \frac{A_{I}}{A_{0}}\, .
\end{align}

Finally, the mass, which in the extremal case is not an independent parameter,
is given by $M=-\dot{U}(0)$. In terms of the integration constants:
\begin{equation}
\label{eq:massformulad5-1}
M = \tilde{H}^{I}(A) B_{I}\, .   
\end{equation}
\noindent
This expression can be rewritten as
\begin{equation}
\label{eq:massformulad5-2}
M = h^{I}_{\infty}B_{I} \equiv \mathcal{Z}_{\rm e}(\varphi_{\infty},B)\, ,  
\end{equation}
\noindent
by analogy with $\mathcal{Z}_{\rm e}(\varphi_{\infty},q)$, the central charge
of the theory. In general, the constants $B_{I}$ will not be equal to the
electric charges $q_{I}$ and the {\it fake central charge} $\mathcal{Z}_{\rm
  e}(\varphi_{\infty},B)$ will differ from the genuine supesymmetric central
charge.

It is also possible and useful to derive generic expressions for the values of
the scalars on the horizon ($\rho\rightarrow \infty$) and for the the hyperarea
$\mathcal{A}$ of the horizon (or the Bekenstein-Hawking entropy $S =
\mathcal{A}/4$) using the homogeneity properties of $\mathsf{W}$:
\begin{align}
\varphi_{I\, {\rm h}} 
& =
\frac{B_{I}}{B_{0}}\, ,
\\
\label{eq:AWB}
\frac{\mathcal{A}}{2\pi^{2}}
& = \frac{\mathsf{W}(B)}{2}\, .
\end{align}

Let us now study the equations of motion. First of all, notice that $B_{I}=\pm
q_{I}$ always solves all the equations. These solutions may not be physically
acceptable for certain sign choices, depending on the range of values that the
scalars can take and the particular model.

The equation for $U$ (\ref{eq:d5Ueq}) is automatically satisfied when the
$H_{I}$ are harmonic. The near-horizon limit of the Hamiltonian constraint
(\ref{eq:hamiltoniand5}) reads
\begin{equation}
\label{eq:16052002}
\partial^{I}\partial^{J}\log\mathsf{W}(B) q_{I}q_{J}
\; =\; 
\partial^{I}\partial^{J}\log\mathsf{W}(B) B_{I}B_{J}
\; =\; -3/2\, ,
\end{equation}
\noindent
where the last step follows from homogeneity.  The first term is equal to
$\frac{3}{2} [\mathsf{W}(B)]^{-4/3} V_{\rm bh}(H,q)\bigr\rvert_{\rm h}$, where
$V_{\rm bh}(H,q)\bigr\rvert_{\rm h} =V_{\rm bh}(B,q)$ is the value of the
black-hole potential as a function of the $H_{I}$, evaluated on the
horizon. Then, using eq.~(\ref{eq:AWB}) we find that the entropy is given by
the value of the black-hole potential on the horizon
\cite{Ferrara:1997tw,Meessen:2011bd}
\begin{equation}
\label{eq:AvsVh}
\frac{\mathcal{A}}{2\pi^{2}}
=
\left[-V_{\rm bh}(B,q)\right]^{3/4}\, .  
\end{equation}
\noindent
Similarly, in the near-horizon limit of the equations of motion we find
that\footnote{As a function of the $H$-variables the black-hole potential is
  scale-invariant and, therefore this equation determines the integration
  constants $B_{I}$ up to a common normalization factor which is, on the other
  hand, irrelevant to determine the entropy or the values of the scalars on
  the horizon. The normalization if fixed by the condition
  eq.~(\ref{eq:16052002}).}
\begin{equation}
\label{eq:dVh=0}
\partial^{K}V_{\rm bh}(B,q)=0\, ,  
\end{equation}
\noindent
so the coefficients of $\rho$ in the harmonic functions are those that
extremize the black-hole potential as a function of the $H_{I}$. From this
result we can recover the known fact that the values of the scalars on the
horizon are those that extremize the black-hole potential as a function of the
scalars, using the fact that the black-hole potential is homogeneous of degree
zero in the $H_{I}$.

The asymptotic limit ($\rho \rightarrow 0$) of the Hamiltonian constraint gives
\begin{equation}
\tfrac{3}{2} M^{2} 
-\tfrac{1}{3}  \partial^{I}\partial^{J}\log\mathsf{W}(A) 
B_{I}B_{J} + V_{\rm bh}(A,q) =0\, . 
\end{equation}
\noindent
Comparing this expression with the general Bogomol'nyi bound
\cite{Ferrara:1997tw,Meessen:2011bd}
\begin{equation}
\label{eq:BBound}
M^{2}  + \tfrac{2}{3}g_{xy}(\phi_{\infty}) \Sigma^{x}\Sigma^{y}
 + V_{\rm bh}(\phi_{\infty},q)  = 0\, ,  
\end{equation}
\noindent
we get 
\begin{equation}
\tfrac{1}{2} M^{2} 
-\tfrac{1}{3}  \partial^{I}\partial^{J}\log\mathsf{W}(A) 
B_{I}B_{J} 
=  
\tfrac{2}{3}g_{xy}(\varphi_{\infty}) \Sigma^{x}\Sigma^{y}\, .
\end{equation}
\noindent
These last two equations could be useful insofar as the the scalar charges
$\Sigma^{i}=\Sigma^{i}(\varphi_{\infty},q)$ were known, which is never the case
until the full solution is known.

%%%%%%%%%%%%%%%%%%%%%%%%%%%%%%%%%%%%%%%%%%%%%%%%%%%%%%%%%%%%%%%%%%%%%%
\subsection{First-order flow equations for extremal black holes}
\label{sec:FirstOrder}
%%%%
Following ref.~\cite{Ortin:2011vm} it is easy to derive first-order equations
for extremal (supersymmetric and non-supersymmetric) black holes for which the
$H_{I}(\rho)$ are harmonic functions of the form given in
eq.~(\ref{eq:harmonic}), where the integration constants $B_{I}$ extremize the
black-hole potential as a function of the $H_{I}$, according to
eq.~(\ref{eq:dVh=0}).\footnote{The question remains whether the $H_{I}$ must
  be harmonic for all extremal black holes in all models. In section
  \ref{sec-STU} we give a proof for static, spherically symmetric black holes
  of the $STU$ model.}  We must insist in the fact that, due to the scale
invariance of $V_{\rm bh}(H)$, the extremal values $B_{I}$ are defined only up
to an overall multiplicative constant; this constant is, however, fixed by
eq.~(\ref{eq:16052002})

We want to derive differential flow equations for the metric function $U$ and
for the scalars $\phi^{x}$ (not necessarily parametrized as in
eq.~(\ref{eq:scalardef})). By virtue of eq.~(\ref{eq:30}) and the definition of
the variables $H_{I}$ we can write
\begin{equation}
d e^{-U}  = d(h^{I}h_{I}e^{-U})  = dh^{I} h_{I}e^{-U} + h^{I} d(h_{I}e^{-U}) 
    = h^{I} d(h_{I}e^{-U})  = h^{I} dH_{I}\, , 
\end{equation}
Using then the harmonicity of the $H$'s, eq.~(\ref{eq:harmonic}), we arrive at
\begin{equation}
\label{eq:d5UFlow}
\frac{de^{-U}}{d\rho} = \mathcal{Z}_{\rm e}(\phi,B)\, .
\end{equation}
Note that the above equation is given in terms of $\mathcal{Z}_{\rm e}(\phi
,B)$ and not in terms of the theory's supersymmetric central charge
$\mathcal{Z}_{\rm e}(\phi,q)$.

Similarly we can write
\begin{equation}
\begin{split}
d\phi^{x} & = h^{Ix}h_{Iy}d\phi^{y} 
  = -\sqrt{3}h^{Ix}\partial_{y}h_{I}d\phi^{y}
  = -\sqrt{3}h^{Ix} d h_{I} \\
 & = -\sqrt{3}h^{Ix} d (e^{U} e^{-U}h_{I})
  = -\sqrt{3} e^{U} h^{Ix} d H_{I}\, ,
\end{split}
\end{equation}
which after a renewed call to the harmonicity of the $H$'s gives
\begin{equation}
\label{eq:d5PhiFlow}
\frac{d\phi^{x}}{d\rho} =
 -3 e^{U}\partial^{x}\mathcal{Z}_{\rm e}(\phi,B)\, .
\end{equation}
The fixed point of the flow, $\phi_{\rm h}(B)$, determined by the {\it fake
  charges} $B_{I}$ through the extremization of the black hole potential,
eq.~(\ref{eq:dVh=0}), is commonly called an attractor.

The flow equations (\ref{eq:d5UFlow}) and (\ref{eq:d5PhiFlow}) imply
second-order equations which are identical to the equations of motion of the
5-dimensional effective FGK action \cite{Meessen:2011bd}, {\em i.e.\/}\
eqs.~(\ref{eq:39}, \ref{eq:39b}), but with the scalar charges replaced in the
black-hole potential by the constants $B_{I}$:
\begin{align}
\ddot{U} +e^{2U}V_{\rm bh}(\phi,B) & = 0\, ,\\
\ddot{\phi}^{x} +\Gamma_{yz}{}^{x}\dot{\phi}^{y}\dot{\phi}^{z}
+\tfrac{3}{2}e^{2U}V_{\rm bh}(\phi,B) & = 0\, ,
\end{align}
where we introduced the {\em fake black hole potential}
\begin{equation}
\label{eq:fakeBHpot}
V_{\rm bh}(\phi,B)  \equiv  
   -\mathit{a}^{IJ}\, B_{I}B_{J} 
 =  
 -\mathcal{Z}_{\rm e}^{2}(\phi,B)
 - 3\,\partial_{x}\mathcal{Z}_{\rm e}(\phi,B)\partial^{x}\mathcal{Z}_{\rm e}(\phi,B)\, .
\end{equation}
This means that the equations of motion will be satisfied for all these
configurations if and only if the fake black-hole potential is equal to the
true one:
\begin{equation}
\label{eq:Effe1}
    V_{\rm bh}(\phi,B)\;  =\;  V_{\rm bh}(\phi,q)\, .  
\end{equation}

Observe that this equation is, up to an overall factor, nothing but the
Hamiltonian constraint (\ref{eq:hamilextremeansatz}). Consequently, in the
extremal case, if one finds values of $B_{I}$ (we could call them attractor
values, by transfering the notion from $\phi_{\rm h}(B)$) that extremize the
(genuine) black hole potential and satisfy the Hamiltonian constraint, then one
has a solution of all the equations of motion.
\par
The derivation of flow equations for extremal black holes presented here
differs from their previous presentation \cite{kn:Ceresole,kn:Perz,kn:Perz2} in two
aspects: Firstly, no specific form of the relation between the fake and actual
charges is assumed here, instead the derivation is based on the assumption of
harmonicity of the variables $H_{I}$.\footnote{In four dimensions a
  generalization of first-order equations that makes neither assumption was
  made in \cite{Galli:2010mg}.} Secondly, expressing the black hole potential
directly by harmonic functions makes it possible to determine the fake charges
by extremization.
\par

%%%%%%%%%%%%%%%%%%%%%%%%%%%%%%%%%%%%%%%%%%%%%%%%%%%%%%%%%%%%%%%%%%%%%%
%%%%%%%%%%%%%%%%%%%%%%%%%%%%%%%%%%%%%%%%%%%%%%%%%%%%%%%%%%%%%%%%%%%%%%
\subsection{Non-extremal black holes}
\label{sec-d5ansatz}
%%%%
In the simple model studied in ref.~\cite{Meessen:2011bd} it was found that, as
in the 4-dimensional case considered in ref.~\cite{Galli:2011fq}, the
non-extremal black-hole solutions of that model as functions of the variables
$H_{I}(\rho)$ are identical to the extremal ones. The difference is that, now,
the functions $H_{I}(\rho)$ are no longer harmonic ({\em i.e.\/}~linear in
$\rho$) but have the general form
\begin{equation}
\label{eq:ansatz}
H_{I}(\rho)
 = A_{I}\cosh( \mathcal{B}\rho)  + 
 \frac{B_{I}}{\mathcal{B}}\sinh( \mathcal{B}\rho)\, ,   
\end{equation}
\noindent
for some integration constants $A_{I}$ and $B_{I}$ that a priori could be
different from those in the extremal ansatz (\ref{eq:harmonic}) and have to be
determined by solving the equations of motion and by imposing the normalization
of the physical fields at spatial infinity.

It is interesting to see whether the non-extremal black-hole solutions of more
general models also share this generic form.  We start by imposing the
asymptotic boundary conditions on the fields, using the generic definition for
the physical scalars eq.~(\ref{eq:scalardef}). It is easy to see that, again,
asymptotic flatness implies eq.~(\ref{eq:asympflatd5}) and that the integration
constants $A_{I}$ are given by eq.~(\ref{eq:AIvarphi}), so they are actually
the same as in the extremal ansatz. Therefore we only have to determine the
$B_{I}$ plus the non-extremality parameter $\mathcal{B}$ by imposing the
equations of motion.

We will also need the definition of the mass, which is given again by
eqs.~(\ref{eq:massformulad5-1}) and (\ref{eq:massformulad5-2}), and the
expressions for the horizon hyperarea and for the values of the scalars on the
outer and inner horizon horizon, which are now
\begin{align}
\label{eq:varphihD}
\varphi_{I}^{\pm} 
& =
\frac{\mathit{B}_{I}^{\pm}}{\mathit{B}_{0}^{\pm}}\, ,
\\
\label{eq:AWBnonext}
\frac{\mathcal{A}_{\pm}}{2\pi^{2}}
& =
\mathsf{W}(\mathit{B}^{\pm})/2\, .
\end{align}
\noindent
where we have defined the shifted coefficients (``dressed charges''
\cite{Mohaupt:2010fk})
\begin{equation}
\label{eq:shiftedB}
\mathit{B}_{I}^{\pm} \equiv B_{I} \pm \mathcal{B}A_{I}\, .
\end{equation}

Now, first of all, observe that the functions of eq.~(\ref{eq:ansatz}) satisfy
\begin{equation}
\ddot{H}_{I}= \mathcal{B}^{2}H_{I}\, ,  
\end{equation}
\noindent
and then, substituting in eq.~(\ref{eq:d5Ueq}) and using the homogeneity
properties of $\mathsf{W}$, we find that it is identically
satisfied. Substituting into the Hamiltonian constraint and the equations of
motion and using the same properties we can rewrite them in the
form
\begin{align}
\label{eq:hamiltoniand5reduced}
\partial^{I}\partial^{J}
\log\mathsf{W}(H) 
\left( B_{I}^{-}B_{J}^{+}-q_{I}q_{J}\right)
& =
0\, ,
\\
\label{eq:eomd5reduced}
\partial^{K}\partial^{I}\partial^{J}
\log\mathsf{W}(H) 
\left( B_{I}^{-}B_{J}^{+}-q_{I}q_{J}\right)
& =
0\, .
\end{align}

In the near-horizon limits, these equations, upon use of the formula
(\ref{eq:AWBnonext}) for the area of the inner and outer horizons, lead to the
following relations
\begin{align}
\label{eq:AvsVhnonext}
\frac{\mathcal{A}_{\pm}}{2\pi^{2}}
& =
\left[
-\left(1\mp\tfrac{4}{3}\mathcal{B}A_{I} \partial^{I}\log{\mathsf{W}}
\right)^{-1}
V_{\rm bh}\right]^{3/4}\!(\mathit{B}^{\pm})\, ,
\\
\label{eq:dV}
\partial^{K}V_{\rm bh}(\mathit{B}^{\pm})
& =
\pm 6\left(\frac{\mathcal{A}_{\pm}}{2\pi^{2}}\right)^{4/3} 
\mathcal{B}A_{I}
\left[
\partial^{I}\partial^{K}\log{\mathsf{W}}
+\frac{2}{3}\partial^{I}\log{\mathsf{W}}
\,\partial^{K}\log{\mathsf{W}}
\right]\!(\mathit{B}^{\pm})
\, ,  
\end{align}
\noindent
which generalize eqs.~(\ref{eq:AvsVh}) and (\ref{eq:dVh=0}) to the non-extremal
case. Since the right-hand side of eq.~(\ref{eq:dV}) does not vanish in general
for non-extremal black holes, we find that, in general, the values of the
scalars on the horizon do not extremize the black-hole potential. In
section~\ref{sec-d5nonextremaldoublyextremal} we are going to study a class of
non-extremal black holes for which the right-hand side of eq.~(\ref{eq:dV})
does vanish, though. We will be able to give the general form of this class of
solutions for any model of $N=2,d=5$ supergravity.

For models with diagonal $\partial^{I}\partial^{J}\log\mathsf{W}$ (which,
given that in $N=2$ supergravity the polynomial $\mathcal{V}$ must be
homogeneous of degree $3$, comprise only two models, (apart from minimal
supergravity): $STU$, discussed in the next section, and $ST^2$) the equations
of motion (\ref{eq:eomd5reduced}) can be solved by \cite{Mohaupt:2010fk}
\begin{equation}
\label{eq:BIdiagonal}
B_{I} = \pm\sqrt{q_{I}^{2}+\mathcal{B}^2A_{I}^{2}} \quad \text{(no summation),}
\end{equation}
\noindent
which completely determines the dressed charges, and thus the values of scalars
on both horizons, in terms of physical charges, the asymptotic values of the
scalars, eq.~(\ref{eq:Avarphiinfty}) and the non-extremality parameter:
\begin{equation}
\mathit{B}_{I}^{-} = \pm\sqrt{q_{I}^2 + (\mathcal{B}A_{0}\varphi_{I\infty})^2} - \mathcal{B}A_{0}\varphi_{I\infty}\, , \qquad
\mathit{B}_{I}^{+} = \pm\sqrt{q_{I}^2 + (\mathcal{B}A_{0}\varphi_{I\infty})^2} + \mathcal{B}A_{0}\varphi_{I\infty}\, ,
\end{equation}
\noindent
These expressions reduce to ($\pm$ absolute values of) the actual charges when
$\mathcal{B} \to 0$. As expected, due to the dependence on $\varphi_{I\infty}$
there is no attractor mechanism in the proper sense, but from
$B_{I}^{-}B_{I}^{+} = q_{I}^{2}$ we can make an new observation that the
extremal attractor value of a scalar is the geometric mean of the non-extremal
horizon values:
\begin{equation}
(\varphi^{x}_{\rm h})^2 = \varphi^{x}_{-}\varphi^{x}_{+} \quad \text{(no summation).}
\end{equation}

%%%%%%%%%%%%%%%%%%%%%%%%%%%%%%%%%%%%%%%%%%%%%%%%%%%%%%%%%%%%%%%%%%%%%%
\subsection{First-order flow equations for non-extremal black holes}
\label{sec:1stNonExt}
%%%%
The derivation leading to eq.~(\ref{eq:d5UFlow}) can be followed
straightforwardly with a small variation: instead of the coordinate $\rho$, we
need a new coordinate $\hat{\rho}$ which is defined by\footnote{This choice,
  like the parametrization of $H_{I}$, is not unique. One could, for instance,
  take $\hat{\rho} = e^{-2\mathcal{B}\rho}$. An advantage of the $\tanh$
  parametrization is that the asymptotic values of scalars are still governed
  only by $A_I$, as in the extremal case.}
\begin{equation}
  \label{eq:15}
  \hat{\rho}  \equiv \frac{\sinh( \mathcal{B}\rho)}{\mathcal{B}\cosh( \mathcal{B}\rho)}\, ,
  \qquad\mbox{so that}\qquad 
  \cosh(\mathcal{B}\rho) = \frac{1}{\sqrt{ 1-\mathcal{B}\hat{\rho}^{2}\ }} \equiv f(\hat{\rho}) \, ,
\end{equation}
which means that the ansatz for $H_{I}$ in eq.~(\ref{eq:ansatz}) now becomes
the ``almost extremal form''
\begin{equation}
  \label{eq:16}
  H_{I} = f(\hat{\rho})\left( A_{I}+B_{I}\hat{\rho}\right) = f(\hat{\rho})\,\hat{H}_{I}\, .
\end{equation}
Taking into account the above expression in the derivation leading to
eq.~(\ref{eq:d5UFlow}) we find that
\begin{equation}
  \label{eq:14}
  \frac{\partial e^{-\hat{U}}}{\partial\hat{\rho}} = 
      \mathcal{Z}_{\rm e}(\phi,B) \, ,
\end{equation}
where we introduced $\hat{U}=U+\log (f)$. The hatted variables still satisfy
$e^{\hat{U}}\hat{H}_{I}=e^{U}H_{I}=h_{I}$.
\par
The analog of eq.~(\ref{eq:d5PhiFlow}) can be seen to be 
\begin{equation}
  \label{eq:17}
  \frac{\partial\phi^{x}}{\partial\hat{\rho}} = -3\,e^{\hat{U}}
         \partial^{x}\mathcal{Z}_{\rm e}(\phi,B)\, .
\end{equation}
Since eqs.~(\ref{eq:14}) and (\ref{eq:15}) have the same functional form as
eqs.~(\ref{eq:d5UFlow}) and (\ref{eq:d5PhiFlow}), we immediately see that they
lead to the FGK equations of motion, albeit with respect to the new coordinate
$\hat{\rho}$ and the new function $\hat{U}$,\footnote{ Note that from $U=U(H)$
  and the scaling properties of $\mathsf{W}$, we can see that
  $\hat{U}=U(\hat{H})$.  }  {\em i.e.\/}
\begin{align}
  \label{eq:18}
  \partial^{2}_{\hat{\rho}}\hat{U}& = -e^{2\hat{U}}V_{\rm bh}(\phi,B) \, , \\
  \partial^{2}_{\hat{\rho}}\phi^{x} + \Gamma_{yz}{}^{x}\partial_{\hat{\rho}}\phi^{y}\partial_{\hat{\rho}}\phi^{z} & =
     -\tfrac{3}{2}e^{2\hat{U}}\partial^{x}V_{\rm bh}(\phi,B) \, .
\end{align}
One can then ask oneself what the equivalent of eq.~(\ref{eq:Effe1}) is. To
this end we shall rewrite the FGK-equation for $U$ in the
$\hat{\rho}$-coordinate and use eq.~(\ref{eq:14}) to get rid of a term linear
in $\partial_{\hat{\rho}}U$:
\begin{equation}
  \label{eq:19}
  -\mathcal{B}^{2} + 2\mathcal{B}^{2}\hat{\rho}\ e^{\hat{U}}\mathcal{Z}_{\rm e}(\phi,B) =
       e^{2\hat{U}}\!\left(f^{-2}\,V_{\rm bh}(\phi,B) - V_{\rm bh}(\phi,q)\right) .
\end{equation}
This equation is the Hamiltonian constraint written in the new coordinates
$\hat{\rho}$. To see this we need
\begin{equation}
  \label{eq:22}
  e^{-U}\mathcal{Z}_{\rm e}(\phi,B) = \tfrac{2}{3}\,B_{I}\partial^{I}\log(\mathsf{W})\,.
\end{equation}
Performing the same operation on the FGK-equation for the scalar fields, we
find that after using eq.~(\ref{eq:17})
\begin{equation}
  \label{eq:21}
   4\mathcal{B}^{2}\,\hat{\rho}\,e^{\hat{U}}\,\partial^{x}\mathcal{Z}_{\rm e}(\phi,B) = e^{2\hat{U}}\!\left(
           f^{-2}\,\partial^{x}V_{\rm bh}(\phi,B) - \partial^{x}V_{\rm bh}(\phi,q)
     \right) .
\end{equation}
The extra factor of 2 on the left-hand side of the above equation compared to
eq.~(\ref{eq:19}) is surprising, but correct; indeed, differentiating
eq.~(\ref{eq:19}) with respect to $\hat{\rho}$ and using the flow equations
(\ref{eq:14}) and (\ref{eq:17}) we find that
\begin{equation}
  \label{eq:23}
  0 = \partial_{\hat{\rho}}\phi^{y}\ g_{yx}\left[
                     4\mathcal{B}^{2}\hat{\rho}\,e^{\hat{U}}\partial^{x}\mathcal{Z}_{\rm e} -
                     e^{2\hat{U}}\!\left( f^{-2}\,\partial^{x}V_{\rm bh}(B) - \partial^{x}V_{\rm bh}(q)\right)
              \right] .
\end{equation}
This implies that eq.~(\ref{eq:19}) is a constant if eq.~(\ref{eq:21}) is
satisfied, whence we can evaluate it at spatial infinity, {\em i.e.\/}\ $\rho
=0$ and also $\hat{\rho}=0$. This gives
\begin{equation}
  \label{eq:20}
  V_{\rm bh}(\phi_{\infty},q) -  V_{\rm bh}(\phi_{\infty}, B) = \mathcal{B}^{2} e^{-2U_{\infty}} = \mathcal{B}^{2}
\end{equation}
by asymptotic flatness.
\par
The flow (\ref{eq:17}) terminates at the horizon ($\hat{\rho} \to
1/\mathcal{B}$). In the extremal case ($\hat{\rho} \to \infty$) or in the
non-extremal case with constant scalars, since
$\partial_{\hat{\rho}}\phi^{x}\bigr\rvert_{\rm h}=0$, the horizon value of the
scalars will be determined by the location of the fixed point (attractor)
$\partial_{x}\mathcal{Z}_{\rm e}(B) = 0$. For a generic non-extremal solution
the horizon value will be attained in a finite $\hat{\rho}$, before a fixed
point is reached.\footnote{ This in effect is the argument given in
  ref.~\cite{Kallosh:2006bt} as to why the attractor mechanism cannot work for
  non-extremal black holes; other arguments are given in
  ref.~\cite{Goldstein:2005hq}. If we were to extend $\hat{\rho}$ beyond
  $1/\mathcal{B}$ to infinity, the values of scalars would be again determined
  by the ratios of $B_{I}$ (this occurs between the horizons
  \cite{Mohaupt:2010fk}), the values of $B_{I}$, however, now depend on the
  asymptotic boundary conditions.}  We can still evaluate the relevant
equations at the horizon to find
\begin{align}
  \label{eq:31}
  -\mathcal{B}^{2}  + 2\mathcal{B}\,e^{\hat{U}_{\rm h}}\mathcal{Z}_{\rm e}(\phi_{\rm h},B) & = -e^{2\hat{U}_{\rm h}}V_{\rm bh}(\phi_{\rm h},q) \, ,\\
    4\mathcal{B}\,e^{-\hat{U}_{\rm h}}\partial_{x}\mathcal{Z}_{\rm e}(\phi_{\rm h},B) & = -\partial_{x}V_{\rm bh}(\phi_{\rm h},q) \, .
\end{align}

%%%%%%%%%%%%%%%%%%%%%%%%%%%%%%%%%%%%%%%%%%%%%%%%%%%%%%%%%%%%%%%%%%%%%%
%%%%%%%%%%%%%%%%%%%%%%%%%%%%%%%%%%%%%%%%%%%%%%%%%%%%%%%%%%%%%%%%%%%%%%

%%%%%%%%%%%%%%%%%%%%%%%%%%%%%%%%%%%%%%%%%%%%%%%%%%%%%%%%%%%%%%%%%%%%%%
%%%%%%%%%%%%%%%%%%%%%%%%%%%%%%%%%%%%%%%%%%%%%%%%%%%%%%%%%%%%%%%%%%%%%%
%%%%%%%%%%%%%%%%%%%%%%%%%%%%%%%%%%%%%%%%%%%%%%%%%%%%%%%%%%%%%%%%%%%%%%
%%%%%%%%%%%%%%%%%%%%%%%%%%%%%%%%%%%%%%%%%%%%%%%%%%%%%%%%%%%%%%%%%%%%%%

\section{Example black-hole solutions}
\label{sec-d5examples}

In this section we illustrate the use of the H-FGK formalism with some
examples, two of which have supersymmetric and non-supersymmetric attractors
and flat directions. The third one is a generic class of solutions whose main
characteristic is that the physical scalars are constants, and as such are a
generalization of the doubly extremal black holes.

%%%%%%%%%%%%%%%%%%%%%%%%%%%%%%%%%%%%%%%%%%%%%%%%%%%%%%%%%%%%%%%%%%%%%%
%%%%%%%%%%%%%%%%%%%%%%%%%%%%%%%%%%%%%%%%%%%%%%%%%%%%%%%%%%%%%%%%%%%%%%

\subsection{Non-extremal black holes with constant scalars}
\label{sec-d5nonextremaldoublyextremal}

When all the scalar fields of an extremal black-hole solution are constant, it
is known as a doubly extremal black hole. It is natural to consider its
non-extremal generalizations, {\em i.e.\/}~non-extremal black holes with
constant scalars. For the general ansatz (\ref{eq:ansatz}) and the generic parametrization
(\ref{eq:scalardef}) of the physical scalars, this condition requires that
\begin{equation}
\varphi_{I}=
\frac{A_{I}}{A_{0}}  
=
\frac{B_{I}}{B_{0}}
=
\frac{\mathit{B}_{I}^{\pm}}{\mathit{B}_{0}^{\pm}}\, ,
\end{equation}
\noindent
and we can write
\begin{equation}
\label{eq:DExtrBH1}
H_{I} = A_{I} H\, ,
\qquad
H  \equiv \cosh(\mathcal{B}\rho)  + \frac{B_{0}}{A_{0}\mathcal{B}}
\sinh(\mathcal{B}\rho)\, ,
\end{equation}
\noindent
so we have
\begin{equation}
\mathsf{W}(H) = 2H^{3/2}\, .  
\end{equation}
\noindent
The metric is, as expected, that of the 5-dimensional Reissner-Nordstr\"om
black hole in all cases.

The only integration constants that need to be found are $B_{0}$ and
$\mathcal{B}$.  It is convenient to introduce in the problem the mass
parameter, given by eq.~(\ref{eq:massformulad5-1}). In this case, it is just
\begin{equation}
\label{eq:DExtrBHM}
M =   \frac{B_{0}}{A_{0}}\, .
\end{equation}
\noindent
Then the Hamiltonian constraint and the equations of motion take the form
\begin{align}
\partial^{I}\partial^{J}
\log\mathsf{W} (A)
\left[ (M^{2}-\mathcal{B}^{2})A_{I}A_{J}-q_{I}q_{J}\right]
& =
0\, ,
\\
\partial^{K}\partial^{I}\partial^{J}
\log\mathsf{W} (A)
\left[ (M^{2}-\mathcal{B}^{2})A_{I}A_{J}-q_{I}q_{J}\right]
& =
0\, ,
\end{align}
\noindent
and are solved if 
\begin{align}
\label{eq:Bfordoubly}
\mathcal{B}^{2} 
-M^{2} -V_{\rm bh}(A,q)
& =
0\, ,
\\
\label{eq:AttrBforDoubly}
\partial^{K}V_{\rm bh}(A,q)
& =
0\, .  
\end{align}

The first equation is just the general Bogomol'nyi bound for constant scalars
(vanishing scalar charges) and the second, owing to the scale invariance of
$V_{\rm bh}(H)$ tells us that the scalars are not affected by the
non-extremality parameter and everywhere take the values that extremize the
black-hole potential, which are the same as in the extremal case with the same
electric charges. The value of the black-hole potential (in particular, on the
horizon) is also the same as in the extremal case with the same electric
charges. Notice that this implies that $V_{\rm bh}(A,q)$ is a function of the
charges $q$ only.

Using this information in eq.~(\ref{eq:AvsVhnonext}) and invoking the
properties of $\mathsf{W}$, we also obtain an expression that relates the
entropy to the entropy of the extremal black hole with the same electric
charges:
\begin{equation}
\label{eq:AvsVhnonextdoublyext}
\frac{\mathcal{A}_{\pm}}{2\pi^{2}}
=
\left(
- \frac{M \pm \mathcal{B}}{M \mp \mathcal{B}}\,
V_{\rm bh}(\mathit{B}^{\pm})\right)^{3/4} ,
\end{equation}
\noindent
Since $V_{\rm bh}(\mathit{B}^{\pm})=V_{\rm bh}(A)$, combining this expression
with the Bogomol'nyi bound we get the well-known formula
\begin{equation}
\frac{\mathcal{A}_{\pm}}{2\pi^{2}}
=
\left(M \pm \mathcal{B}\right)^{3/2}\, ,  
\end{equation}
\noindent
which also admits the suggestive expression
\begin{equation}
\frac{\mathcal{A}_{\pm}}{2\pi^{2}}
=
\mathcal{Z}_{\rm e}(\varphi_{\infty},\mathit{B}^{\pm})^{3/2}\, ,    
\end{equation}
\noindent
and leads to the suggestive relation
\begin{equation}
\frac{\mathcal{A}_{+}}{2\pi^{2}}\frac{\mathcal{A}_{-}}{2\pi^{2}}  
=
(M^{2}-\mathcal{B}^{2})^{3/2}
=
[-V_{\rm bh}(A,q)]^{3/2}
=
\left(\frac{\mathcal{A}_{\rm ext}}{2\pi^{2}}\right)^{2}\, ,
\end{equation}
\noindent
where, as we stressed above, $V_{\rm bh}(A,q)$ is moduli-independent and
$\mathcal{A}_{\rm ext}$ is hyperarea of the extremal black hole with the same
charges. We will refer to this property in what follows as the
\textit{geometrical mean property}\footnote{A proof for the charged, rotating,
  asymptotically-flat or anti-de-Sitter black-hole solutions of a wide class
  of theories (which does not include those we are considering here) has been
  given in \cite{Cvetic:2010mn}. Earlier, less general results, were found in
  \cite{Cvetic:1996kv,Larsen:1997ge,Cvetic:1997uw,Cvetic:1997xv,Cvetic:2009jn}. A
  related result valid for horizons of arbitrary topology has been recently
  found in \cite{Castro:2012av}.}.

Summarizing, the solutions of this class, for any model, are obtained by
finding first the values (determined up to a common factor) of the
$\mathit{B}_{I}^{\pm}$ that extremize the potential $\partial^{K}V_{\rm
  bh}(\mathit{B}^{\pm})=0$. The scalars are then given by
$\varphi_{I}=\mathit{B}_{I}^{\pm}/\mathit{B}_{0}^{\pm}$, which, through
eq.~(\ref{eq:asympconstantsd5-1}), dictates the constants $A_{I}$ for these
values of the scalars. The non-extremality parameter is established by
eq.~(\ref{eq:Bfordoubly}), the metric function is $e^{-U/2}= H$ with $H$ as in
eq.~(\ref{eq:DExtrBH1}), and the mass of this black hole is found from
eq.~(\ref{eq:DExtrBHM}).

%%%%%%%%%%%%%%%%%%%%%%%%%%%%%%%%%%%%%%%%%%%%%%%%%%%%%%%%%%%%%%%%%%%%%%
\subsubsection{Constant-scalar black holes from the flow equations}
%%%%%%%%%%%
As one can see from eq.~(\ref{eq:17}), constant scalars around a black hole satisfy
\begin{equation}
  \label{eq:24}
  \partial_{x}\mathcal{Z}_{\rm e}(B) = 0 \qquad \Rightarrow \qquad
  \partial_{x}V_{\rm bh}(\phi ,B) = 0\, .
\end{equation}
We can then use eq.~(\ref{eq:14}) to obtain
\begin{equation}
  e^{-\hat{U}} = 1 + \mathcal{Z}_{\rm e}(B)\hat{\rho} \, ,
\end{equation}
where we already imposed asymptotic Minkowskianity. It is also evident that the
mass of the solution is given by $M=\mathcal{Z}_{\rm e}(B)$. Plugging the
conditions in eq.~(\ref{eq:24}) into eq.~(\ref{eq:21}) we see that
\begin{equation}
  \label{eq:25}
  \partial_{x}V_{\rm bh}(\phi ,q) = 0 \, ,
\end{equation}
which is the analog of eq.~(\ref{eq:AttrBforDoubly}).
Eq.~(\ref{eq:Bfordoubly}) can be derived immediately from the Hamiltonian
constraint (\ref{eq:20}).
\par
Eq.~(\ref{eq:24}) says that, in terms of the fake charges, the constant scalars
of a non-supersymmetric solution have the same form as the scalars of the
supersymmetric extremal solution in terms of the real charges, whereas
eq.~(\ref{eq:25}) fixes the scalars directly in terms of $q$'s.

%%%%%%%%%%%%%%%%%%%%%%%%%%%%%%%%%%%%%%%%%%%%%%%%%%%%%%%%%%%%%%%%%%%%%%
%%%%%%%%%%%%%%%%%%%%%%%%%%%%%%%%%%%%%%%%%%%%%%%%%%%%%%%%%%%%%%%%%%%%%%
%%%%%%%%%%%%%%%%%%%%%%%%%%%%%%%%%%%%%%%%%%%%%%%%%%%%%%%%%%%%%%%%%%%%%%
%%%%%%%%%%%%%%%%%%%%%%%%%%%%%%%%%%%%%%%%%%%%%%%%%%%%%%%%%%%%%%%%%%%%%%

\subsection{The \texorpdfstring{$STU$}{STU} model revisited}
\label{sec-STU}

The $STU$ model in five dimensions is defined by 
\begin{equation}
\mathcal{V}(h^{\cdot}) = h^{0}h^{1}h^{2}=1\, .  
\end{equation}
\noindent
The corresponding unconstrained function in the H-FGK formalism is 
\begin{equation}
\mathsf{V}(\tilde{H}) = \tilde{H}^{0}\tilde{H}^{1}\tilde{H}^{2} \, .
\end{equation}
\noindent
The tilded variables are given in terms of the untilded ones by 
\begin{equation}
\tilde{H}^{0} =\sqrt{\frac{3H_{1}H_{2}}{H_{0}}}\, ,
\qquad
\tilde{H}^{1,2}= \frac{3H_{2,1}}{\tilde{H}^{0}}\, ,
\end{equation}
\noindent
so
\begin{equation}
\mathsf{W}(H) = 2\,\mathsf{V}(H)= 2\sqrt{3^{3}H_{0}H_{1}H_{2}}\, .  
\end{equation}
This potential contains all the information that we need to find and construct
all the black-hole solutions of the model.

Due to the special form of $\mathsf{W}$, the equations of motion
(\ref{eq:eomsd5}) are completely separated and read
\begin{equation}
\label{eq:STUeom}
     H_{I}\ddot{H}_{I} - \dot{H}_{I}^{2} + q_{I}^{2} = 0 \quad \text{(no summation)} \, .
\end{equation}
These equations can be integrated explicitly, with the general solution being
of the form
\begin{equation}
\label{eq:KiloUtrechtTango}
  H_{I} = a_{I}\cosh( \varepsilon_{I}\rho) + b_{I}\sinh( \varepsilon_{I}\rho) \, .
\end{equation}
The Hamiltonian constraint, eq.~(\ref{eq:hamiltoniand5}), then imposes the
condition $\sum_{I}\varepsilon_{I}^{2}=3\mathcal{B}^{2}$. A further constraint
arises due to the fact that we are interested in building black holes as
expressed by eq.~(\ref{eq:33}), which implies that
$\sum_{I}\varepsilon_{I}=3\mathcal{B}$. In addition, we should also have scalar
fields that are regular on the horizon. We can do this by imposing the
condition that the $\varphi_{I}$ be regular there, which also ensures the
regularity of the physical scalars $\phi^{x}$ on the horizon. Clearly this
means that $\varepsilon_{0} =\varepsilon_{1} =\varepsilon_{2} = \mathcal{B}$,
reducing the general solution to the ansatz (\ref{eq:ansatz}). In other words,
in the $STU$ model, all black-hole solutions of the type considered here must
be described by this ansatz.

%%%%%%%%%%%%%%%%%%%%%%%%%%%%%%%%%%%%%%%%%%%%%%%%%%%%%%%%%%%%%%%%%%%%%%
%%%%%%%%%%%%%%%%%%%%%%%%%%%%%%%%%%%%%%%%%%%%%%%%%%%%%%%%%%%%%%%%%%%%%%
%%%%%%%%%%%%%%%%%%%%%%%%%%%%%%%%%%%%%%%%%%%%%%%%%%%%%%%%%%%%%%%%%%%%%%
%%%%%%%%%%%%%%%%%%%%%%%%%%%%%%%%%%%%%%%%%%%%%%%%%%%%%%%%%%%%%%%%%%%%%%
%%%%%%%%%%%%%%%%%%%%%%%%%%%%%%%%%%%%%%%%%%%%%%%%%%%%%%%%%%%%%%%%%%%%%%

\subsubsection{Extremal solutions}

In the limit $\mathcal{B}\rightarrow 0$ the ansatz (\ref{eq:ansatz}) reduces
to eq.~(\ref{eq:harmonic}) and it follows from the argument above that all
extremal solutions to the $STU$ model (at least of the kind we are analyzing)
will be described by harmonic functions.

To determine their coefficients, let us first analyze the critical points of
the black-hole potential $V_{\rm bh}(H,q)$, which for the $STU$ model takes the
form
\begin{equation}
  V_{\rm bh}(H,q)  
  =
  \tfrac{2}{3} \left(\mathsf{W}/2 \right)^{4/3} \partial^{I}\partial^{J}
  \log{\mathsf{W}}\,q_{I}q_{J}
= -3 (H_{0}H_{1}H_{2})^{2/3}\sum_{I} \left(\frac{q_{I}}{H_{I}}\right)^{2}\, .
\end{equation}
\noindent
The equations $\partial^{K}V_{\rm  bh}(H,q)\bigr\rvert_{\rm h} = 0$ are solved by
\begin{equation}
(B_{I})^{2} =\alpha^{2}(q_{I})^{2}\, ,
\end{equation}
\noindent
where $\alpha$ is an arbitrary constant, resulting from the scale invariance of
the black-hole potential as a function of the $H$'s. This constant does not
affect the attractor points of the physical scalars, which are given by
quotients of $H$'s.

The above solutions correspond to   
\begin{equation}
B_{I}= s_{I}q_{I}\, ,  
\end{equation}
\noindent
where the signs $s_{I} = \pm 1$ can, in principle, be chosen at will and are
independent of the electric charges. Each choice of signs corresponds to a
different kind of solution that may or may not (we will carefully look into
this point) describe several signs of the charges. The reality and
regularity of the metric and scalar fields will impose certain restrictions on
the possible signs, though. First of all, the reality of $\mathsf{W}(B)$
requires that
\begin{equation}
\label{eq:constraintsssqqq}
\beta \operatorname{sgn}(q_{0}) \operatorname{sgn}(q_{1}) \operatorname{sgn}(q_{2})
=+1\, ,
\qquad
\beta \equiv s_{0}s_{1}s_{3}\, .
\end{equation}
\noindent
The attractor values for the physical scalars, chosen as in
eq.~(\ref{eq:scalardef}) for $x=1,2$, on the horizon are
\begin{equation}
\label{eq:STUattractors}
\varphi_{x\, {\rm h}} = s_{0}s_{x} q_{x}/q_{0}\, .  
\end{equation}
\noindent
In terms of these scalars the sections are given by 
\begin{align}
h_{0} & = \frac{1}{3(\varphi_{1}\varphi_{2})^{1/3}}\, ,
& 
h_{1} & = \frac{\varphi_{1}}{3(\varphi_{1}\varphi_{2})^{1/3}}\, ,
& 
h_{2} & = \frac{\varphi_{2}}{3(\varphi_{1}\varphi_{2})^{1/3}}\, ,
 \\
\label{eq:hvarphi}
h^{0} & = (\varphi_{1}\varphi_{2})^{1/3}\, , 
&
h^{1} & = \frac{(\varphi_{1}\varphi_{2})^{1/3}}{\varphi_{1}}\, ,   
& 
h^{2} & = \frac{(\varphi_{1}\varphi_{2})^{1/3}}{\varphi_{2}}\, .   
\end{align}

Observe that the scalars can be positive or negative but not zero. This means
that the theory has four branches\footnote{The scalar manifold has to be
  covered by four coordinate patches.} that can be labeled by the four possible
combinations of the two signs of the scalars: $\sigma_{x} \equiv
\operatorname{sgn}(\varphi_{x})$. In terms of unconstrained scalars $\phi_{x}$
(customarily called for this model $S$ and $T$) we would have $\varphi_{x} \sim
\sigma_{x} e^{\phi_{x}}$. Since, in a regular solution for a given branch
$\sigma_{1},\sigma_{2}$, the above scalars will have the same sign everywhere
and in particular on the horizon, from eq.~(\ref{eq:STUattractors}) we find
that admissible regular extremal solutions in the branch $\sigma_{x}$ satisfy
besides eq.~(\ref{eq:constraintsssqqq}) also
\begin{equation}
\label{eq:constraintsqssq}
s_{0}\, s_{x} = \sigma_{x} \operatorname{sgn}(q_{0}) \operatorname{sgn}(q_{x}) \, . 
\end{equation}
\noindent
This condition ensures that $\operatorname{sgn}(B_{I})
=\operatorname{sgn}(A_{I})$ for all $I$, so the functions $H_{I}$ do not vanish
for any positive value of the radial coordinate $\rho$, which is another
condition for regularity of the solution. In any given branch $\sigma_{x}$, for
any of the 8 possible choices of signs of the charges, there is always a choice
of $s_{I}$ that allows us to have a regular extremal solution.

To find out which of these solutions are supersymmetric in each branch, we need
to extremize the central charge, which is given by
\begin{equation}
\mathcal{Z}_{\rm e}(\varphi,q) =   
(\varphi_{1}\varphi_{2})^{-2/3} 
\left( q_{0}\varphi_{1}\varphi_{2} +q_{1}\varphi_{2}
  +q_{2}\varphi_{1}\right)\, .
\end{equation}
\noindent
The extrema correspond to the horizon values
\begin{equation}
  \varphi^{\rm SUSY}_{x\, {\rm h}}= q_{x}/q_{0}\, .  
\end{equation}
\noindent
Comparing with eq.~(\ref{eq:STUattractors}) we find that only the cases
\begin{equation}
 \operatorname{sgn}(q_{0})\, \operatorname{sgn}(q_{x}) = \sigma_{x}\, ,  
\end{equation}
\noindent
are supersymmetric in the branch $\sigma_{x}$. This corresponds to two possible
sign configurations for each of the four branches so all charge configurations
are supersymmetric in some branch, as in the case considered in
ref.~\cite{Meessen:2011bd}.

\begin{table}
  \centering
\label{tab:STUbranch+-}
\begin{tabular}{ccccccc}
\hline  
$ \operatorname{sgn}(q_{0})$ & $ \operatorname{sgn}(q_{1})$ &  
$\operatorname{sgn}(q_{3})$ & $s_{0}$ & $s_{1}$ & $s_{2}$ & SUSY \\
\hline  
$+$ & $+$ & $+$ & $-$ & $-$ & $+$ & no  \\
$+$ & $+$ & $-$ & $-$ & $-$ & $-$ & yes \\
$+$ & $-$ & $+$ & $-$ & $+$ & $+$ & no  \\
$+$ & $-$ & $-$ & $-$ & $+$ & $-$ & no  \\
$-$ & $+$ & $+$ & $+$ & $-$ & $+$ & no  \\
$-$ & $+$ & $-$ & $+$ & $-$ & $-$ & no  \\
$-$ & $-$ & $+$ & $+$ & $+$ & $+$ & yes \\
$-$ & $-$ & $-$ & $+$ & $+$ & $-$ & no  \\
\hline  
\end{tabular}
\caption{In the last three columns of this table we give the choices of signs $s_{I}$
  necessary to have regular solutions for each of the combinations of signs of
  the electric charges given in the first three columns in the branch 
  $\sigma_{1}=1,\, \sigma_{2}=-1$. The supersymmetric configurations of this 
  branch are those in the second and seventh rows.}
\end{table}

Combining the choices of $s_{I}$ with the signs of the charges, all the
solutions can be written in a unified way in terms of the harmonic functions
\begin{equation}
H_{I} = \frac{\operatorname{sgn}(q_{0})\, \operatorname{sgn}(q_{1})\,
  \operatorname{sgn}(q_{2})}{\operatorname{sgn}(q_{I})}
\left(
\frac{4^{1/3}}{3} 
\left\lvert
\frac{\varphi_{I\, \infty}}{(\varphi_{1\, \infty\,}\varphi_{2\, \infty})^{1/3}}
\right\rvert
+|q_{I}|\rho
\right)\, ,  
\end{equation}
\noindent
which are manifestly non-vanishing for positive values of $\rho$ and are valid
for all four branches and all eight combinations of the signs of the charges.
Furthermore, in all cases and for all choices of the signs of the charges
(contrary to what is stated in ref.~\cite{Mohaupt:2011aa}), the metric function
$e^{-\frac{3}{2}U}= \mathsf{W}(H)/2$ is real and regular. The entropy, given by
eq.~(\ref{eq:AWB}), takes the explicit form
\begin{equation}
\frac{\mathcal{A}}{2\pi^{2}} = \tfrac{1}{2}\sqrt{3^{3}|q_{0}q_{1}q_{2}|}\, ,  
\end{equation}
\noindent
and the mass, given by eq.~(\ref{eq:massformulad5-2}) becomes
\begin{equation}
M = (\varphi_{1\, \infty\,}\varphi_{2\, \infty})^{-2/3}
\Bigl( 
|\varphi_{1\, \infty\,}\varphi_{2\, \infty\,}q_{0}| 
+|\varphi_{2\, \infty\,}q_{1}| 
+|\varphi_{1\, \infty\,}q_{2}| 
\Bigr)\, ,  
\end{equation}
\noindent
which is always positive. It coincides with the supergravity central charge at
infinity only for certain signs of the charges that depend on the branch
considered, as we have explained before.

%%%%%%%%%%%%%%%%%%%%%%%%%%%%%%%%%%%%%%%%%%%%%%%%%%%%%%%%%%%%%%%%%%%%%%
%%%%%%%%%%%%%%%%%%%%%%%%%%%%%%%%%%%%%%%%%%%%%%%%%%%%%%%%%%%%%%%%%%%%%%
%%%%%%%%%%%%%%%%%%%%%%%%%%%%%%%%%%%%%%%%%%%%%%%%%%%%%%%%%%%%%%%%%%%%%%
%%%%%%%%%%%%%%%%%%%%%%%%%%%%%%%%%%%%%%%%%%%%%%%%%%%%%%%%%%%%%%%%%%%%%%
%%%%%%%%%%%%%%%%%%%%%%%%%%%%%%%%%%%%%%%%%%%%%%%%%%%%%%%%%%%%%%%%%%%%%%

\subsubsection{Non-extremal solutions}

For the present model the ansatz (\ref{eq:ansatz}) is the general solution to
the equations of motion (\ref{eq:eomd5reduced}) or (\ref{eq:STUeom}) and they
reduce to the following relation between the parameters:
\begin{equation}
\label{eq:BIBAIqI}
B_{I}^{2} =\mathcal{B}^{2}A_{I}^{2}+q_{I}^{2}\, .
\end{equation}
\noindent
Since the integration constants $A$ are, according to the general arguments,
given by
\begin{equation}
A_{I} = \frac{4^{1/3}}{3}\frac{\varphi_{I\, \infty}}{(\varphi_{1\,
    \infty\,}\varphi_{2\, \infty})^{1/3}}\, ,  
\end{equation}
\noindent
the above equations immediately give the complete solution
\begin{equation}
B_{I} = s_{I} \sqrt{q_{I}^{2}+\mathcal{B}^{2}A_{I}^{2}}\, ,  
\end{equation}
\noindent
where we have to choose the signs $s_{I}$ so that the functions $H_{I}$ do not
vanish for any value of $\rho>0$, {\em i.e.\/}\ so that
$\operatorname{sgn}(B_{I}) =\operatorname{sgn}(A_{I})$ (assuming
$\mathcal{B}>0$):
\begin{equation}
s_{I} =  \frac{\operatorname{sgn}(\varphi_{0\, \infty})\, \operatorname{sgn}(\varphi_{1\, \infty})\,
  \operatorname{sgn}(\varphi_{2\, \infty})}{\operatorname{sgn}(\varphi_{I\,
    \infty})}\, . 
\end{equation}

The regularity of the metric translates into manifest positivity of the mass,
given by eq.~(\ref{eq:massformulad5-2}) and, explicitly, by
\begin{equation}
\begin{split}
M ={}& \sqrt{\left[(\varphi_{1\, \infty\,}\varphi_{2\,
      \infty})^{1/3}q_{0}\right]^{2} 
+\mathcal{B}^{2}\left[(\varphi_{1\, \infty\,}\varphi_{2\,
    \infty})^{1/3}A_{0}\right]^{2}}  \\
&+\sqrt{\left[\frac{\varphi_{2\, \infty\,}}{(\varphi_{1\, \infty\,}\varphi_{2\,
      \infty})^{1/3}}q_{1}\right]^{2} 
+\mathcal{B}^{2}\left[\frac{\varphi_{2\, \infty\,}}{(\varphi_{1\, \infty\,}\varphi_{2\,
      \infty})^{1/3}}A_{1}\right]^{2}}  \\
&+\sqrt{\left[\frac{\varphi_{1\, \infty\,}}{(\varphi_{1\, \infty\,}\varphi_{2\,
      \infty})^{1/3}}q_{2}\right]^{2} 
+\mathcal{B}^{2}\left[\frac{\varphi_{1\, \infty\,}}{(\varphi_{1\, \infty\,}\varphi_{2\,
      \infty})^{1/3}}A_{2}\right]^{2}} \, .
\end{split}
\end{equation}

The non-extremality parameter can be solved in terms of the mass, asymptotic
values of the scalars and charges by solving a quartic algebraic equation in
$\mathcal{B}^{2}$, but the expression is too complicated to be useful.

The hyperareas of the horizons, given by the general formula
(\ref{eq:AWBnonext}), take the form
\begin{equation}
\frac{\mathcal{A}_{\pm}}{2\pi^{2}} = \frac{1}{2}
\left[3^{3} 
\prod_{I}\left( \sqrt{q_{I}^{2}+\mathcal{B}^{2}A_{I}^{2}}\pm\mathcal{B}|A_{I}| \right)
\right]^{1/2} \, .
\end{equation}
\noindent
We can see explicitly that not only the values of scalars, as mentioned
earlier, but also the entropies $S_{\pm} = \mathcal{A}_{\pm}/4$ satisfy the
geometric mean property:
\begin{equation}
S_{-} S_{+} = S^2 \, ,
\end{equation}
where the mean value is that of the extremal black hole.

%%%%%%%%%%%%%%%%%%%%%%%%%%%%%%%%%%%%%%%%%%%%%%%%%%%%%%%%%%%%%%%%%%%%%%
%%%%%%%%%%%%%%%%%%%%%%%%%%%%%%%%%%%%%%%%%%%%%%%%%%%%%%%%%%%%%%%%%%%%%%
%%%%%%%%%%%%%%%%%%%%%%%%%%%%%%%%%%%%%%%%%%%%%%%%%%%%%%%%%%%%%%%%%%%%%%
%%%%%%%%%%%%%%%%%%%%%%%%%%%%%%%%%%%%%%%%%%%%%%%%%%%%%%%%%%%%%%%%%%%%%%
%%%%%%%%%%%%%%%%%%%%%%%%%%%%%%%%%%%%%%%%%%%%%%%%%%%%%%%%%%%%%%%%%%%%%%

\subsection{Models of the generic Jordan family}
\label{sec-Jordan}

The models of the reducible Jordan sequence are defined by
\begin{equation}
\mathcal{V}(h^{\cdot}) = h^{0}\eta_{ij}h^{i}h^{j} =1\, ,  
\qquad
i=1,\dotsc,n\, ,
\end{equation}
\noindent
where $(\eta_{ij}) = \operatorname{diag}(-+\cdots+)$ and the associated
potential in the H-FGK formalism takes the form
\begin{equation}
\mathsf{V}(\tilde{H}) = \tilde{H}^{0} \tilde{H}^{2}\, , 
\end{equation}
\noindent
where we have defined
\begin{equation}
\tilde{H} ^{2} \equiv \tilde{H}^{i}\tilde{H}_{i} \equiv  
\eta_{ij}\tilde{H}^{i}\tilde{H}^{j} \equiv \tilde{H} \cdot \tilde{H}\, . 
\end{equation}

The relation between tilded and untilded variables is
\begin{equation}
\tilde{H}^{0} =\frac{1}{2}\sqrt{\frac{3H^{2}}{H_{0}}}\, ,  
\qquad
\tilde{H}^{i} = H^{i}\sqrt{\frac{3H^{0}}{H^{2}}}\, ,  
\end{equation}
\noindent
so
\begin{equation}
\mathsf{W}(H) = 2\mathsf{V}(H) = \sqrt{3^{3}H_{0}H^{2}}\, ,  
\end{equation}
\noindent
The non-vanishing components of the Hessian of $\log{\mathsf{W}}$ are
\begin{equation}
\partial^{0}  \partial^{0} \log{\mathsf{W}} = 
-\frac{1}{2 H_{0}^{2}}\, ,
\qquad
\partial^{i}  \partial^{j} \log{\mathsf{W}} = 
\frac{\eta^{ij} H^{2} -2H^{i}H^{j}}{(H^{2})^{2}}\, .
\end{equation}

%%%%%%%%%%%%%%%%%%%%%%%%%%%%%%%%%%%%%%%%%%%%%%%%%%%%%%%%%%%%%%%%%%%%%%
%%%%%%%%%%%%%%%%%%%%%%%%%%%%%%%%%%%%%%%%%%%%%%%%%%%%%%%%%%%%%%%%%%%%%%
%%%%%%%%%%%%%%%%%%%%%%%%%%%%%%%%%%%%%%%%%%%%%%%%%%%%%%%%%%%%%%%%%%%%%%
%%%%%%%%%%%%%%%%%%%%%%%%%%%%%%%%%%%%%%%%%%%%%%%%%%%%%%%%%%%%%%%%%%%%%%
%%%%%%%%%%%%%%%%%%%%%%%%%%%%%%%%%%%%%%%%%%%%%%%%%%%%%%%%%%%%%%%%%%%%%%

\subsubsection{Extremal solutions}

There are two kinds of critical loci of the black-hole potential: the discrete
points
\begin{align}
B_{i} & = s q_{i}\, ,
\quad
s^{2}=+1\, , 
\\
\frac{B^{2}}{B^{2}_{0}} -\frac{q^{2}}{q^{2}_{0}} & = 0\, ,
\end{align}
\noindent
including those that correspond to supersymmetric black holes, and the
$(n-1)$-dimensional space described by the constraint
\begin{align}
\label{eq:Bq=0}
B\cdot q & = 0\, ,
\\
\label{eq:secondcondition}
\frac{B^{2}}{B^{2}_{0}} +\frac{q^{2}}{q^{2}_{0}} & = 0\, ,
\end{align}
\noindent
which gives rise to non-supersymmetric solutions. We focus on the latter since
we expect the scalars on the horizon to depend on the values of the scalars at
infinity.

The constraint eq.~(\ref{eq:Bq=0}) is solved in a general way by
\begin{equation}
B_{i} = \alpha \left[ (C\cdot q) q_{i} -q^{2} C_{i}\right] ,   
\end{equation}
\noindent
for some constants $C_{i}$ that are defined only up to shifts proportional to
the charges $q_{i}$ and up to a normalization constant $\alpha$.  The limit of
the Hamiltonian constraint (\ref{eq:hamilextremeansatz}) on the horizon is
solved by
\begin{equation}
B^{2} = -q^{2}\, ,  
\end{equation}
\noindent
and then the second condition eq.~(\ref{eq:secondcondition}) determines the
integration constants $B_{0}$
\begin{equation}
B_{0}^{2} = q_{0}^{2} \qquad \Rightarrow \qquad B_{0}= s_{0}q_{0}\, , \quad
s_{0}^{2}=+1\, .  
\end{equation}
\noindent
Plugging the general solution for $B_{i}$ into $B^{2}= -q^{2}$ we find that
\begin{equation}
\alpha^{2} = \left[ (C\cdot q)^{2} -C^{2} q^{2} \right]^{-1}\qquad \Rightarrow \qquad 
B_{i} = s\frac{(C\cdot q) q_{i} -q^{2} C_{i}}{\sqrt{ (C\cdot q)^{2} -C^{2}
    q^{2}}}\, ,\quad s^{2}=+1\, ,
\end{equation}
\noindent
so the coefficients $B_{i}$ have a highly non-linear dependence on the charges,
something that could make us think that the $H_{i}$ might be highly non-linear
functions of harmonic functions. Evidently, we have assumed from the onset the
harmonicity of these variables and our challenge is to prove that the ansatz
solves all the equations of motion for this moduli space of non-supersymmetric
attractors.

The asymptotic limit of the Hamiltonian constraint
(\ref{eq:hamilextremeansatz}) is solved if the constants $C_{i}$ are
proportional to the $A_{i}$, whose value is known. We take the proportionality
constant to be $1$ and then it becomes just a matter of calculation to see that
eq.~(\ref{eq:hamilextremeansatz}) is satisfied everywhere. The $K=0$ component
of the equations of motion (\ref{eq:eomextremeansatz}) is trivially satisfied
and, again, it is a matter of calculation to check that the $K=k$ components,
which have the form
\begin{equation}
\left( 3 \eta^{(ij}H^{k)}H^{2} -4H^{i}H^{j}H^{k} \right)
( B_{i}B_{j}-q_{i}q_{j})=0\, ,
\end{equation}
\noindent
are also identically satisfied for 
\begin{equation}
B_{i} = s\frac{(A\cdot q) q_{i} -q^{2} A_{i}}{\sqrt{ (A\cdot q)^{2} -A^{2}
    q^{2}}}\, ,\quad s^{2}=+1\, .  
\end{equation}

%%%%%%%%%%%%%%%%%%%%%%%%%%%%%%%%%%%%%%%%%%%%%%%%%%%%%%%%%%%%%%%%%%%%%%
%%%%%%%%%%%%%%%%%%%%%%%%%%%%%%%%%%%%%%%%%%%%%%%%%%%%%%%%%%%%%%%%%%%%%%
%%%%%%%%%%%%%%%%%%%%%%%%%%%%%%%%%%%%%%%%%%%%%%%%%%%%%%%%%%%%%%%%%%%%%%
%%%%%%%%%%%%%%%%%%%%%%%%%%%%%%%%%%%%%%%%%%%%%%%%%%%%%%%%%%%%%%%%%%%%%%
%%%%%%%%%%%%%%%%%%%%%%%%%%%%%%%%%%%%%%%%%%%%%%%%%%%%%%%%%%%%%%%%%%%%%%

\subsubsection{Non-extremal solutions}

It is not too difficult to extend the extremal solutions to the non-extremal
regime using the ansatz (\ref{eq:ansatz}). For that purpose it is convenient to
introduce the mass parameter. According to the general expression
(\ref{eq:massformulad5-2}), it is given by
\begin{equation}
M = \frac{2}{3} \left(\frac{B_{0}}{2A_{0}} +\frac{A\cdot B}{A^{2}} \right) ,   
\end{equation}
\noindent
and we can use this formula to express  the combination $A\cdot B$ as
\begin{equation}
A\cdot B = \frac{A^{2}}{2} \left( 3 M - \frac{B_{0}}{A_{0}}\right).  
\end{equation}
\noindent
All the terms in the right-hand side of this expression are known in terms of
physical parameters except for $B_{0}$, but this constant can be found by
solving the $K=0$ equation of motion:
\begin{equation}
\label{eq:B0BA0q0}
B_{0} = s_{0} \sqrt{q_{0}^{2} +\mathcal{B}^{2}A^{2}_{0}}\, ,  
\end{equation}
\noindent
so 
\begin{equation}
\label{eq:AdotB}
A\cdot B = \frac{A^{2}}{2} \left( 3 M 
-\frac{s_{0}}{A_{0}}\sqrt{q_{0}^{2} +\mathcal{B}^{2}A^{2}_{0}}\right) .  
\end{equation}
\noindent
In order to keep the expressions as simple as possible, we will not replace
$A\cdot B$ by its above value in what follows.

The $K=k$ equations of motion 
\begin{equation}
\left( 3 \eta^{(ij}H^{k)}H^{2} -4H^{i}H^{j}H^{k} \right)
\left( B_{i}B_{j}-\mathcal{B}^{2}A_{i}A_{j}-q_{i}q_{j}\right)=0\, ,
\end{equation}
\noindent
upon use of the Hamiltonian constraint
\begin{equation}
H^{2} (B^{2} -\mathcal{B}^{2} A^{2} -q^{2})
-2 \left[(H\cdot B)^{2} -\mathcal{B}^{2}(H\cdot A)^{2} -(H\cdot q)^{2}
\right] =0\, , 
\end{equation}
\noindent
can be expanded in a finite number of powers of $\tanh{\mathcal{B}\rho}$ and,
requiring that all the coefficients vanish, we get two equations:
\begin{align}
A^{k} (B^{2} -\mathcal{B}^{2} A^{2} -q^{2}) 
-2  \left[B^{k} (A\cdot B) -\mathcal{B}^{2} A^{k} A^{2} -q^{k} (A\cdot q)
\right] & = 0\, ,\\
B^{k} (B^{2} -\mathcal{B}^{2} A^{2} -q^{2}) 
-2  \left[B^{k} B^{2} -\mathcal{B}^{2} A^{k} (A\cdot B) -q^{k} (B\cdot q)
\right] & = 0\, .
\end{align}
\noindent
It can be checked that these two equations imply the Hamiltonian constraint,
therefore it is enough to solve only them. These equations contain two unknown
combinations of integration constants: $B^{2}$ and $B\cdot q$, which can be
found by multiplying the above equations by $A_{k}, B_{k}$ or $q_{k}$. We get
\begin{align}
B^{2} 
& =
\mathcal{B}^{2} A^{2} +q^{2} +\frac{2}{A^{2}}
\left[(A\cdot B)^{2} -\mathcal{B}^{2}(A^{2})^{2} -(A\cdot q)^{2}
\right] ,\\
B\cdot q
& =
\frac{A\cdot B}{(A\cdot q) A^{2}}
\left[q^{2}A^{2}+ (A\cdot B)^{2} -\mathcal{B}^{2}(A^{2})^{2} -(A\cdot q)^{2}
\right] .  
\end{align}
 
Substituting these expressions in the two equations above we obtain two
different equations for $B^{k}$ in terms of the known objects $(A\cdot B),
\mathcal{B}, A^{k}, A^{2}, (A\cdot q)$:
\begin{align}
B^{k}
& =
\frac{[(A\cdot B)^{2} - (A\cdot q)^{2}] A^{k} +A^{2} (A\cdot q)
  q^{k}}{A^{2}(A\cdot B)}\, ,\\
B^{k}
& =
\frac{(A\cdot B)
\left\{A^{2} (A\cdot q) \mathcal{B}^{2} A^{k} 
+[q^{2}A^{2}+ (A\cdot B)^{2} -\mathcal{B}^{2}(A^{2})^{2} -(A\cdot q)^{2}]
q^{k}\right\} }{(A\cdot q)\left[q^{2}A^{2}+ (A\cdot B)^{2} -(A\cdot q)^{2}
\right]}\, .
\end{align}
\noindent
These two solutions must be equal and one can see that this happens when the
following condition is satisfied:
\begin{equation}
(A^{2})^{2}(A\cdot B)^{2}\mathcal{B}^{2}
= [(A\cdot B)^{2} - (A\cdot q)^{2}]^{2}
+A^{2}q^{2} [(A\cdot B)^{2} - (A\cdot q)^{2}]\, .  
\end{equation}
\noindent
This condition, on account of eq.~(\ref{eq:AdotB}), is an equation that
involves $M,\mathcal{B}^{2},A_{I}$ and $q_{I}$ and, in principle, may be used
to express the non-extremality parameter as
$\mathcal{B}^{2}(M,\varphi_{x\,\infty},q_{I})$.

%%%%%%%%%%%%%%%%%%%%%%%%%%%%%%%%%%%%%%%%%%%%%%%%%%%%%%%%%%%%%%%%%%%%%%
%%%%%%%%%%%%%%%%%%%%%%%%%%%%%%%%%%%%%%%%%%%%%%%%%%%%%%%%%%%%%%%%%%%%%%
%%%%%%%%%%%%%%%%%%%%%%%%%%%%%%%%%%%%%%%%%%%%%%%%%%%%%%%%%%%%%%%%%%%%%%
%%%%%%%%%%%%%%%%%%%%%%%%%%%%%%%%%%%%%%%%%%%%%%%%%%%%%%%%%%%%%%%%%%%%%%
%%%%%%%%%%%%%%%%%%%%%%%%%%%%%%%%%%%%%%%%%%%%%%%%%%%%%%%%%%%%%%%%%%%%%%
\section{H-FGK formalism for black-string solutions}
\label{sec:StringFGK}
%%%%%%%%%%%%%%%%%%%%%%%%%%%%%%%%%%%%%%%%%%%%%%%%%%%%%%%%%%%%%%%%%%%%%%
In this section we will develop a formalism analogous to the one in section
\ref{sec-d5}, but for obtaining string-like solutions; the derivation follows
similar lines, the only difference being the identification of the new
variables.  Indeed, as one can see from
refs.~\cite{Chamseddine:1999qs,Gauntlett:2004qy,Bellorin:2007yp}, the
seed-functions for supersymmetric string-like solutions are not related to the
$h_{I}$ as in the black hole, but rather to the $h^{I}$.  As such, the
formalism to be developed and illustrated in this section will be based on new
variables $K^{I}$ and $\tilde{K}_{I}$, ($I=0,\ldots ,n$), which we define by
\begin{equation}
  \label{eq:changev}
  K^{I} \equiv e^{-U} h^{I}(\phi ) \, .
\end{equation}
By substituting this change of variables into the fundamental constraint of
real special geometry and defining
\begin{equation}
  \label{eq:2}
  \mathsf{V}(K) \equiv C_{IJK}K^{I}K^{J}K^{K}\, , \qquad \mbox{we find that} \qquad
  e^{-3U} = \mathsf{V}(K)\, .
\end{equation}
We can then introduce the dual variables $\tilde{K}_{I}$ by
\begin{equation}
  \label{eq:3}
  \tilde{K}_{I} = e^{-2U} h_{I}(\phi ) \qquad \mbox{or equivalently} \qquad
  \tilde{K}_{I} = \tfrac{1}{3}\partial_{I}\mathsf{V}(K) \, ,
\end{equation}
but they will be used sparingly in this section.
\par
The FGK formalism for black holes can be generalized to the case of branes
\cite{Martin:2012bi}.  The generic metric for 5-dimensional black-string
solutions is
\begin{equation}
\label{eq:172}
       ds^{2} = e^{U(\rho)-\mathcal{B}\rho}dt^{2}  -e^{U(\rho)+\mathcal{B}\rho}dy^{2}
       -e^{-2U(\rho)} \left( \frac{\mathcal{B}^4}{\sinh^4(\mathcal{B}\rho)}d\rho^{2} + \frac{\mathcal{B}^2}{\sinh^2(\mathcal{B}\rho)}d\Omega^{2}_{(2)} \right) ,    
\end{equation}
\noindent
where $d\Omega^{2}_{(2)}$ denotes the round metric on the 2-sphere. Observe
that the function $U$ in the above metric must satisfy
$\lim_{\rho\rightarrow\infty}(U+\mathcal{B}\rho) =0$ in order for the metric to
describe a black string with a horizon located at $\rho\rightarrow\infty$. If
this condition is met, the near-horizon geometry for $\mathcal{B}\neq 0$ is a
2-dimensional Rindler space times $\mathbb{R}\times S^{2}$; in the extremal
case, $\mathcal{B}=0$, the near-horizon geometry is $aDS_{3}\times S^{2}$ as
usual.
\par
We consider purely ``magnetic'' black string solutions, meaning that we take
$F^{I}\sim\star_{(5)}(dv\wedge dt\wedge dx)$, the implication of which is that
we can safely ignore the Chern-Simons term in the parent $d=5$ supergravity
action and straightforwardly Hodge-dualize the $F^{I}$. The resulting action
reads
\begin{equation}
  \label{eq:168}
  \mathcal{I}_{5} = \int_{5}\sqrt{g}\left( R  + \tfrac{1}{2} \mathit{g}_{xy}\partial_{\mu}\phi^{x}\partial^{\mu}\phi^{y}
                 + \tfrac{1}{2\cdot 3!} a^{IJ}G_{I \mu\nu\kappa}G_{J}^{\mu\nu\kappa}
          \right) ,
\end{equation}
where $G_{I}=dB_{I}$.  The resulting equations of motion for the above action
are
\begin{equation}
  \label{eq:175}
  R_{\mu\nu} = -\tfrac{1}{2}\mathit{g}_{xy}\partial_{\mu}\phi^{x}\partial_{\nu}\phi^{y}
        - \tfrac{1}{4}a^{IJ}\left( G_{I\mu\kappa\lambda}G_{J\nu}{}^{\kappa\lambda} - \tfrac{2}{9}\eta_{\mu\nu}G_{I}\cdot G_{J}\right) .
\end{equation}
Given the ansatz for the metric, the $B$ equation of motion is readily solved
to give
\begin{equation}
  \label{eq:169}
  G_{I} = \sqrt{3}\ e^{2U}a_{IJ}\,\mathit{p}^{J} dv\wedge dt\wedge dx \, ,
\end{equation}
where $\mathit{p}^{I}$ are the string charges and the $\sqrt{3}$ is inserted
for convenience.
\par
By simply substituting the ansatz into eq.~(\ref{eq:175}) we find that
\begin{align}
  \label{eq:176}
  \ddot{U} & = -e^{2U}a_{IJ}\mathit{p}^{I}\mathit{p}^{J} \, , \\
  \mathcal{B}^{2} & = \dot{U}^{2} + \tfrac{1}{3}\mathit{g}_{xy}\dot{\phi}^{x}\dot{\phi}^{y}
                                    - e^{2U}a_{IJ}\mathit{p}^{I}\mathit{p}^{J} \, .
\end{align}
We will not give the equations of motion for the scalars as the result should
be obvious. The resulting FGK action can then be seen with the aid of
eqs.~(\ref{eq:35}) and (\ref{eq:36}) to be
\begin{equation}
\label{eq:effectived52strings}
\mathcal{I}[U,\phi^{x}]   = \int d\rho
    \left(
          \dot{U}^{2}
          + a_{IJ}\dot{h}^{I}\dot{h}^{J}
          - e^{2U}V_{\rm st}
          + \mathcal{B}^{2}
\right) ,
\end{equation}
where we have defined the {\em black string potential} as
\begin{equation}
  \label{eq:4}
  V_{\rm st} \equiv - a_{IJ}\,\mathit{p}^{I}\mathit{p}^{J}
   = -\mathcal{Z}_{\rm m} - 3 \partial_{x}\mathcal{Z}_{\rm m}\partial^{x}\mathcal{Z}_{\rm m} \, ,
\end{equation}
where in the last step we introduced the {\em (magnetic) string central charge}
$\mathcal{Z}_{\rm m}=h_{I}\mathit{p}^{I}$; $\mathcal{B}$ is again a
non-extremality parameter.  In order to obtain the above action in the
$K$-variables, we will need the straightforward identity
\begin{equation}
\label{eq:zooi}
     -3\, e^{2U}a_{IJ} = \partial_{I}\partial_{J}
      \log\mathsf{V} \equiv \mathit{v}_{IJ}\, ,  
\end{equation}
\noindent
which then enables us to write eq.~(\ref{eq:effectived52strings}) as
\begin{equation}
\label{eq:StringHFGKaction}
-3\,\mathcal{I}[K] 
 = \int\! d\rho
       \left(
             \mathit{v}_{IJ}\bigl( 
                     \dot{K}^{I}\dot{K}^{J} + \mathit{p}^{I}\mathit{p}^{J}
            \bigr)
        - 3\mathcal{B}^{2}
\right) .  
\end{equation}
The Hamiltonian constraint can be expressed as
\begin{equation}
  \label{eq:hamiltoniand5strings}
      \mathcal{H} \equiv  
            \mathit{v}_{IJ}
                \bigl( 
                     \dot{K}^{I}\dot{K}^{J}
                      - \mathit{p}^{I}\mathit{p}^{J}
               \bigr)
      + 3\mathcal{B}^{2} = 0
\end{equation}
\noindent
and the equations of motion derived from the effective action are
\begin{equation}
\label{eq:eomsd5strings}
\partial_{I}\mathit{v}_{KL}\bigl(
                   \dot{K}^{K}\dot{K}^{L} - K^{K}\ddot{K}^{L} - \mathit{p}^{K}\mathit{p}^{L}
              \bigr) = 0\, .
\end{equation}
After contraction with $K^{I}$ and some minor manipulations they lead to 
\begin{equation}
  \label{eq:5}
  \ddot{K}^{I}\partial_{I}\log\mathsf{V} = 3\mathcal{B}^{2}\, ,
\end{equation}
which can be written as $\bigl(\ddot{K}_{I} - \mathcal{B}^{2}K^{I}\bigr)\partial_{I}\mathsf{V} = 0$.

The resulting equations are very similar to the ones obtained in the black-hole
case ({\em cf.\/}\ eqs.~(\ref{eq:hamiltoniand5}, \ref{eq:eomsd5},
\ref{eq:d5Ueq})), we thus make the ansatz of the same type as
eq.~(\ref{eq:ansatz}),
\begin{equation}
  \label{eq:6}
  K^{I} = A^{I}\cosh( \mathcal{B}\rho) + \frac{B^{I}}{\mathcal{B}}\sinh(\mathcal{B}\rho)\, ,
\end{equation}
which allows us to write eq.~(\ref{eq:hamiltoniand5strings}) in the more
manageable form
\begin{equation}
  \label{eq:7}
  \mathit{v}_{KL}\left(B^{K}B^{L} - \mathcal{B}^{2}A^{K}A^{L} - \mathit{p}^{K}\mathit{p}^{L}\right) = 0 \, .
\end{equation}
In the same way we can write the equations of motion,
\begin{equation}
  \label{eq:8}
  \left(\partial_{I}\mathit{v}_{KL}\right)
     \left(B^{K}B^{L} - \mathcal{B}^{2}A^{K}A^{L} - \mathit{p}^{K}\mathit{p}^{L}\right) = 0\, ,
\end{equation}
which can be seen as an extremization condition for eq.~(\ref{eq:7}).
\par
In order to make further contact with the supergravity fields, we mimic the
definition of the scalar fields in eq.~(\ref{eq:scalardef}) by defining
\begin{equation}
  \label{eq:9}
  \varphi^{I} \equiv \frac{h^{I}}{h^{0}} = \frac{K^{I}}{K^{0}}\, , \qquad\mbox{so that}\qquad \varphi^{0}\equiv 1 \, .
\end{equation}
As before, we can then fix $A^{I}$ in terms of the asymptotic values of the
scalar fields $\varphi_{\infty}^{I}$ by
\begin{equation}
  \label{eq:10}
  A^{I} = \mathsf{V}(\varphi_{\infty})^{-1/3} \varphi^{I}_{\infty} \, .
\end{equation}
\par
Following ref.~\cite{Martin:2012bi}, we can calculate the string tension to be
\begin{equation}
  \label{eq:41}
  \mathcal{T}_{(1)} = \mathcal{B} + \tfrac{3}{4}\mathit{B}^{I}_{-}\tilde{A}_{I} \, ,
  \qquad\mbox{where}\qquad
  \mathit{B}^{I}_{-} \equiv B^{I} - \mathcal{B}A^{I}
\end{equation}
and we defined $\tilde{A}_{I}=\lim_{\rho\rightarrow 0}\tilde{K}_{I}$, which by
eq.~(\ref{eq:10}) satisfy $A^{I}\tilde{A}_{I}=1$. The values of physical
quantities on the (outer) horizon are given by the shifted components analogous
to the ones defined in eq.~(\ref{eq:shiftedB})
\begin{equation}
  \label{eq:42}
  \mathit{B}^{I}_{+} = B^{I} + \mathcal{B}A^{I} \, ,
\end{equation}
which allows us to express the string's tension as
\begin{equation}
  \label{eq:43}
  \mathcal{T}_{(1)} = \tfrac{3}{4}\mathit{B}^{I}_{+}\tilde{A}_{I} - \tfrac{1}{2}\mathcal{B} \, .
\end{equation}
The temperature of the black string is easily calculated to be
\begin{equation}
  \label{eq:44}
  T_{+} = \frac{4\pi}{\sqrt{2\mathcal{B}}}\,\mathsf{V}^{-1/2}(\mathit{B}_{+})\, ,
\end{equation}
and ref.~\cite{Martin:2012bi}'s entropy density $\mathcal{S}_{+}$ is\footnote{
  $\mathcal{S}_{\pm} \equiv \text{Area}_{\pm}/4$, where $\text{Area}_{\pm}$
  is the area of the 2-sphere in the near-outer (respectively
  near-inner)-horizon geometry.  }
\begin{equation}
  \label{eq:45}
  \mathcal{S}_{+} = \mathsf{V}^{2/3}(\mathit{B}_{+}) \, ,
  \qquad\mbox{whence}\qquad
  \sqrt{2\mathcal{B}} = 4\pi T_{+} \mathcal{S}_{+}^{3/4} \, ,
\end{equation}
if full concordance with the general results obtained in
ref.~\cite{Martin:2012bi}.
\par
The metric (\ref{eq:172}), by an extension of the argument that we gave earlier
for black holes, can cover also the interior of the inner horizon, except that
the r\^{o}les of coordinates $t$ and $y$ become interchanged.\footnote{ It is
  perhaps useful to introduce the coordinate $r$ by $r\cdot
  (1-e^{-2\mathcal{B}\rho})=2\mathcal{B}$, which takes the FGK metric in
  eq.~(\ref{eq:172}) to the standard form with a blackening factor.
   %\textbf{This definition coincides with \cite{Mohaupt:2010fk}'s $\rho^2$. Also, in the past we used $S_{\pm}$ for the outer/inner horizon.}
} 
We can then calculate the temperature and the entropy density on the inner horizon to be
\begin{equation}
  \label{eq:47}
  T_{-} = \frac{4\pi}{\sqrt{2\mathcal{B}}}\, \mathsf{V}^{-1/2}(\mathit{B}_{-}) \, , \qquad
  \mathcal{S}_{-} = \mathsf{V}^{2/3}(\mathit{B}_{-}) \qquad
  \mbox{and} \qquad 
  \sqrt{2\mathcal{B}} = 4\pi T_{-} \mathcal{S}_{-}^{3/4} \, . 
\end{equation}

%%%%%%%%%%%%%%%%%%%%%%%%%%%%%%%%%%%%%%%%%%%%%%%%%%%%%%%%%%%%%%%%%%%%%%
\subsection{Flow equations for black strings}
\label{sec:FlowString}
%%%%
As in section \ref{sec:FirstOrder}, we can derive general flow equations. In
this case we can use $h^{I}=e^{U}K^{I}=e^{\hat{U}}\hat{K}^{I}$, where
$\hat{K}^{I}$ is a function of a new coordinate $\hat{\rho}$ such that
\begin{equation}
  \label{eq:26}
  \frac{\partial \hat{K}^{I}}{\partial \hat{\rho}} = B^{I} \, .
\end{equation}
Using then the completeness and orthogonality relations of real special
geometry, we find that the above equation is equivalent to the following system
of flow equations
\begin{align}
  \label{eq:27}
  \partial_{\hat{\rho}}\hat{U} & = -e^{\hat{U}}\mathcal{Z}_{\rm m}(B) \, ,\\
  \label{eq:27a}
  \partial_{\hat{\rho}}\phi^{x} & = -3\,e^{\hat{U}}\partial^{x}\mathcal{Z}_{\rm m}(B) \, ,
\end{align}
where we defined the {\em fake magnetic (dual) central charge}
$\mathcal{Z}_{\rm m}(B)=h_{I}B^{I}$.  Observe that the above equations are,
{\em mutatis mutandis}, identical to flow equations (\ref{eq:14}, \ref{eq:17})
for the black holes. This means that as long as we are considering the same
kind of ansatz for the seed functions, which is the case, we will find that the
above flow equations will lead to a solution of the FGK equations of motion as
long as
\begin{equation}
  \label{eq:28}
    4\mathcal{B}^{2}\hat{\rho}\,e^{\hat{U}}\partial^{x}\mathcal{Z}_{\rm m}(\phi ,B) = e^{2\hat{U}}\bigl[
           f^{-2}\partial^{x}V_{\rm st}(\phi ,B) - \partial^{x}V_{\rm st}(\phi ,\mathit{p})
     \bigr] \, ,
\end{equation}
and the non-extremality parameter can be obtained from
\begin{equation}
  \label{eq:29}
  V_{\rm st}(\phi_{\infty},\mathit{p}) - V_{\rm st}(\phi_{\infty}, B) = \mathcal{B}^{2} e^{-2U_{\infty}} = \mathcal{B}^{2} \, .
\end{equation}
%%%%%%%%%%%%%%%%%%%%%%%%%%%%%%%%%%%%%%%%%%%%%%%%%%%%%%%%%%%%%%%%%%%%%%
%%%%%%%%%%%%%%%%%%%%%%%%%%%%%%%%%%%%%%%%%%%%%%%%%%%%%%%%%%%%%%%%%%%%%%
\subsection{Non-extremal black strings with constant scalars}
\label{sec:DoubleBlackStrings}
%%%%
As in section~\ref{sec-d5nonextremaldoublyextremal}, we can consider the
non-extremal analog of the doubly extremal string solution, by which we mean a
black string-like solution with constant physical scalars. Using the ansatz
(\ref{eq:6}) and the shorthand notation of eq.~(\ref{eq:42}), we see
immediately that
\begin{equation}
  \label{eq:11}
  \frac{A^{I}}{A^{0}} = \frac{B^{I}}{B^{0}} =
  \frac{\mathit{B}^{I}_{\pm}}{\mathit{B}^{0}_{\pm}} \, .
\end{equation}
The general form of the $K$'s then becomes
\begin{equation}
  \label{eq:1}
  K^{I} = A^{I}\,\mathcal{K}\, , \qquad
  \mathcal{K} = \cosh(\mathcal{B}\rho) + 
    \frac{\mathcal{E}}{\mathcal{B}}\sinh(\mathcal{B}\rho)\, ,
\end{equation}
with a constant of proportionality
$\mathcal{E}=\tfrac{B^{0}}{A^{0}}$, which must be positive in order
for the metric to be well-defined:
\begin{equation}
  \label{eq:12}
  e^{-U} = \left[\mathsf{V}(A)\right]^{1/3}\mathcal{K} = \mathcal{K}\, ,
\end{equation}
where we used asymptotic flatness: $\mathsf{V}(A) = 1$.  In this general case
the tension, eq.~(\ref{eq:43}), and the entropy densities, eqs.~(\ref{eq:45})
and (\ref{eq:47}), can be calculated to be
\begin{equation}
  \label{eq:46}
  \mathcal{T}_{(1)} = \tfrac{1}{4}( 3\mathcal{E}+\mathcal{B})\, ,
  \qquad
  \mathcal{S}_{\pm} = ( \mathcal{E} \pm \mathcal{B})^{2} \, .
\end{equation}
\par
As always, the precise relation between various constants appearing in the
solution, notably $\mathcal{E}$ and $\mathcal{B}$, has to be fixed by the
equations of motion (\ref{eq:8}) and the Hamiltonian constraint
(\ref{eq:8}). For the case at hand they can be recast in the form
\begin{align}
  \label{eq:13}
  \mathcal{E}^{2} + V_{\rm st}(A,p) & = \mathcal{B}^{2}\, , \\
  \label{eq:13a}
  [\partial_{I}V_{\rm st}](A,p) & = 0\, .
\end{align}
The evident similarity to eqs.~(\ref{eq:Bfordoubly}) and
(\ref{eq:AttrBforDoubly}) was to be expected.
%%%%%%%%%%%%%%%%%%%%%%%%%%%%%%%%%%%%%%%%%%%%%%%%%%%%%%%%%%%%%%%%%%%%%%
%%%%%%%%%%%%%%%%%%%%%%%%%%%%%%%%%%%%%%%%%%%%%%%%%%%%%%%%%%%%%%%%%%%%%%
\subsection{Extremal strings}
\label{sec-extremalstrings}
%%%%%%
In this case we are interested in solutions for which $\mathcal{B}=0$.  By
defining
\begin{equation}
  \label{eq:197}
  V_{\rm st}(K,B) = -\mathit{a}_{IJ}B^{I}B^{J} 
  \qquad\mbox{and also}\qquad
  V_{\rm st}(K,\mathit{p}) =
  -\mathit{a}_{IJ}\mathit{p}^{I}\mathit{p}^{J} \, ,
\end{equation}
where the $K$-dependence resides in $\mathit{a}_{IJ}$ through
eq.~(\ref{eq:zooi}), we can see that the equations of motion and the
Hamiltonian constraint can be written as
\begin{equation}
  \label{eq:193}
    \partial_{I}\bigl( 
      V_{\rm st}(K,B) - V_{\rm st}(K,\mathit{p}) \bigr) 
     = 0 \qquad\mbox{and}\qquad
   V_{\rm st}(K,B) = V_{\rm st}(K,\mathit{p}) \, ,
\end{equation}
the former being implied by the latter. 
\par
The above results are nothing new as they also follow from the flow equations,
but it is interesting to evaluate them on the horizon\footnote{These equations
  are analogous to eqs.~(\ref{eq:dVh=0}) and (\ref{eq:16052002}).}:
\begin{equation}
  \label{eq:194}
  \left[\partial_{I}V_{\rm st}\right](B,\mathit{p}) = 0
  \qquad\mbox{and}\qquad
  \mathit{v}_{KL}(B)\,\mathit{p}^{K}\mathit{p}^{L} = -3 \, .
\end{equation}
%%%%%%%%%%%%%%%%%%%%%%%%%%%%%%%%%%%%%%%%%%%%%%%%%%%%%%%%%%%%%%%%%%%%%%
\subsubsection{Extremal strings of the \texorpdfstring{$STU$}{STU} model}
%%%%%%%%%%%%%%%%%%%%%%%%%%%%%%%%%%%%%%%%%%%%%%%%%%%%%%%%%%%%%%%%%%%%%%
%%%%%%%%%%%%%%%%%%%%%%%%%%%%%%%%%%%%%%%%%%%%%%%%%%%%%%%%%%%%%%%%%%%%%%
%%%%%%%%%%%%%%%%%%%%%%%%%%%%%%%%%%%%%%%%%%%%%%%%%%%%%%%%%%%%%%%%%%%%%%
%%%%%%%%%%%%%%%%%%%%%%%%%%%%%%%%%%%%%%%%%%%%%%%%%%%%%%%%%%%%%%%%%%%%%%
The 5-dimensional $STU$ model can be obtained as a consistent truncation of a
6-torus compactification of M-theory, meaning that any 5-dimensional solution
can always be lifted to M-theory. As is well-known, the supersymmetric black
holes derived in sec.~\ref{sec-STU} correspond to the intersection of three
M2-branes, which after a chain of dualities leads to {\em e.g.\/}~a D5-D1-F1
intersection that is used to calculate the microscopic entropy
\cite{Strominger:1996sh}. This identification can also be used to explain the
microscopic origin of the near-extremal black holes \cite{Horowitz:1996ay}.
The uplift of the supersymmetric strings is readily identified with an
intersection of three M5-branes and the general string solutions can be seen to
be deformations of these intersections.
\par
The relevant polynomial reads
\begin{equation}
\label{eq:VSTUstrings}
\mathsf{V}(K)= K^{0} K^{1} K^{2}\, .
\end{equation}
\noindent
which, when comparing it to the $\mathrm{W}(H)$ in sec.~\ref{sec-STU} and
remembering that the Hesse metric is the second derivative of the logarithm of
$\mathsf{V}$, means that the problem is completely analogous to the problem
treated in section \ref{sec-STU}.  It is therefore no great surprise to see
that the Hamiltonian constraint and the equations of motion are given
respectively by
\begin{align}
\label{eq:HSTUstrings}
3\mathcal{B}^{2} & = \sum_{I}\frac{\ddot{K}^I}{K^I}\, ,\\
\label{eq:EqSTUstrings}
0& = K^I \ddot{K}^I-(\dot{K}^I)^2+(p^I)^2 \quad \text{(no summation)}.
\end{align}
The general comments made in sec.~\ref{sec-STU} about the separability of the
equations and their solutions apply also in this case.
\par
%%%%%%%%%%%%%%%%%%%%%%%%%%%%%%%%%%%%%%%%%%%%%%%%%%%%%%%%%%%%%%%%%%%%%%
%%%%%%%%%%%%%%%%%%%%%%%%%%%%%%%%%%%%%%%%%%%%%%%%%%%%%%%%%%%%%%%%%%%%%%
%%%%%%%%%%%%%%%%%%%%%%%%%%%%%%%%%%%%%%%%%%%%%%%%%%%%%%%%%%%%%%%%%%%%%%
%%%%%%%%%%%%%%%%%%%%%%%%%%%%%%%%%%%%%%%%%%%%%%%%%%%%%%%%%%%%%%%%%%%%%%
The extremal solutions, taking into account the branches and the signs, are
given by harmonic functions
\begin{equation}
  \label{eq:48}
  K^{I} = s^{I}\operatorname{sgn}(p^{I}) \left( |A^{I}| + |p^{I}|\rho \right) , 
  \quad (s^{I})^{2} = 1 \, ,
\end{equation}
where the various signs have to satisfy
\begin{equation}
  \label{eq:49}
  1 = s^{0}s^{1}s^{2}\operatorname{sgn}(p^{0}) \operatorname{sgn}(p^{1}) \operatorname{sgn}(p^{2})
  \qquad\mbox{and}\qquad
  s^{0}s^{x} = \sigma^{x} \operatorname{sgn}(p^{0}) \operatorname{sgn}(p^{x}) \, ,
\end{equation}
which are completely analogous to eqs.~(\ref{eq:constraintsssqqq}) and
(\ref{eq:constraintsqssq}).  The supersymmetric extremal solutions, {\em
  i.e.\/}~the ones that extremize the string central charge $\mathcal{Z}_{\rm
  m}(\phi ,p)$, have signs that satisfy
\begin{equation}
  \label{eq:50}
  \operatorname{sgn}(p^{0}) = \sigma^{x} \operatorname{sgn}(p^{x}) \; .
\end{equation}
Table \ref{tab:STUbranch+-} can also be applied to this case, and gives the
possible sign choices for the $(+,-)$ branch, the second and seventh row being
supersymmetric.
\par
The entropy density for the extremal strings is
\begin{equation}
  \label{eq:51}
  \mathcal{S} = \left| p^{0} p^{1} p^{2} \right|^{2/3} .
\end{equation}
The string's tension can be calculated from eq.~(\ref{eq:43}) and reads
\begin{equation}
  \label{eq:52}
  \mathcal{T}_{(1)} = \frac{1}{4|\varphi_{\infty}^{1}\varphi^{2}_{\infty}|^{2/3}} \left(
            |\varphi^{1}_{\infty}\varphi^{2}_{\infty}\mathit{p}^{0}|  +
            |\varphi^{2}_{\infty}\mathit{p}^{1}| +
            |\varphi^{1}_{\infty}\mathit{p}^{2}|
     \right) ,
\end{equation}
when expressed in terms of the asymptotic values of the scalar fields
$\varphi^{x}$.
%%%%%%%%%%%%%%%%%%%%%%%%%%%%%%%%%%%%%%%%%%%%%%%%%%%%%%%%%%%%%%%%%%%%%%
%%%%%%%%%%%%%%%%%%%%%%%%%%%%%%%%%%%%%%%%%%%%%%%%%%%%%%%%%%%%%%%%%%%%%%
%%%%%%%%%%%%%%%%%%%%%%%%%%%%%%%%%%%%%%%%%%%%%%%%%%%%%%%%%%%%%%%%%%%%%%
%%%%%%%%%%%%%%%%%%%%%%%%%%%%%%%%%%%%%%%%%%%%%%%%%%%%%%%%%%%%%%%%%%%%%%
\subsubsection{Extremal strings in the heterotic \texorpdfstring{$STU$}{STU} model}
%%%%%%%%
In this section we consider the extremal string solutions to the $STU$ model
with a correction, leaving the non-extremal ones for future work as they are
much more involved.
\par
The model that we want to consider can be obtained by compatifying heterotic
string theory on $K3\times S^{1}$ and the fundamental polynomial is given by
\cite{Antoniadis:1995vz}
\begin{equation}
  \label{eq:53}
  \mathcal{V}(h^{\cdot}) = \begin{cases}
       h^{0}h^{1}h^{2} + \tfrac{\aleph^{2}}{3}\left(h^{0}\right)^{3}   & \text{for}\quad h^{0}<h^{1}\, , \\[1ex]
      h^{0}h^{1}h^{2} + \tfrac{\aleph^{2}}{3}\left(h^{1}\right)^{3}    & \text{for}\quad h^{0}>h^{1}\, ,
  \end{cases}
\end{equation}
where $\aleph =1$ has been introduced in order to be able to discuss the $STU$
limit $\aleph\rightarrow 0$.  The line $h^{0}=h^{1}$ corresponds to the
selfdual radius of the circle compactification, where extra massless modes
arise; $h^{2}$ corresponds to the 5-dimensional dilaton
\cite{Antoniadis:1995vz}.  We shall restrict ourselves to the case when
$h^{I}>0$, hence also $\varphi^{I} > 0$, and we shall furthermore restrict
ourselves to the wedge of moduli space where $h^{1}>h^{0}$, or alternatively
$\varphi^{1}>1$, in order not to have to deal with solutions that interpolate
between the two wedges.\footnote{ The interesting case of having a solution
  that switches from one wedge to another will not be considered here.  See
  {\em e.g.\/}~\cite{Gaida:1998km} for supersymmetric black hole and string
  solutions that do switch wedges.  }
\par
Let us in passing mention that the BPS black holes based on this model were
obtained by Gaida in ref.~\cite{Gaida:1998pz}, who showed that there is a
quantum constraint on the electric charges. This restriction arises as follows:
by eq.~(\ref{eq:35}) we see that
\begin{equation}
  \label{eq:54}
    3h_{0} = h^{1}h^{2} + \aleph^{2}(h^{0})^{2} \, , \qquad
    3h_{1} = h^{0}h^{2} \, , \qquad 
    3h_{2} = h^{0}h^{1} \, ,
\end{equation}
which can be inverted over the complex numbers to give
\begin{equation}
  \label{eq:55}
  h^{1} = \frac{3h_{2}}{h^{0}} \, , \qquad
  h^{2} = \frac{3h_{1}}{h^{0}} \, , \qquad
  \tfrac{2}{3}\aleph^{2}(h^{0})^{2}  = h_{0} \pm \sqrt{h_{0}^{2} - 4\aleph^{2} h_{1}h_{2} } \, .
\end{equation}
Since the $h$'s must be real,
\begin{equation}
  \label{eq:56}
  (h_{0})^{2} \geq 4\aleph^{2} h_{1}h_{2} 
  \qquad\mbox{or dually:}\qquad
  \left( h^{1}h^{2} - \aleph^{2}(h^{0})^{2} \right)^{2} \geq 0 \, ,
\end{equation}
to which we shall refer to as Gaida's bound and which is a restriction
originating from the well-definedness of the model in real special geometry.
As the restriction must also hold on the horizon, the attractor mechanism
implies Gaida's constraint $q_{0}^{2}\geq 4\aleph^{2} q_{1}q_{2}$
\cite{Gaida:1998pz}.
\par
The supersymmetric solutions can be found easily by extremizing the string
central charge $\mathcal{Z}_{\rm m}(\mathit{p})$.  To that end one would in
principle need a parameterization of the $h$'s in terms of the physical scalars
$\phi^{x}$, but it is advantageous to use the $\varphi^{x}$ as physical scalars
as then the attractor equation
\begin{equation}
  \label{eq:57}
  0 = \left. \frac{\partial\mathcal{Z}_{\rm m}(\mathit{p})}{\partial \varphi^{x} }\right|_{\varphi^{x}_{\rm h}} ,
\end{equation}
becomes readily solvable by seeing that $h^{x}=h^{0} \varphi^{x}$ and
$(h^{0})^{-3}=\mathsf{V}(\varphi^{I})$.  Perhaps surprisingly, this equation
has two solutions, namely:
\begin{itemize}
\item[a)]  The first solution is given by 
  \begin{equation}
    \label{eq:59}
    \varphi^{1}_{\rm h} = \frac{B^{1}}{B^{0}} = \frac{\mathit{p}^{1}}{\mathit{p}^{0}} \, ,
    \qquad
    \varphi^{2}_{\rm h} = \frac{B^{2}}{B^{0}} = \frac{\mathit{p}^{2}}{\mathit{p}^{0}} \, ,
  \end{equation}
  which is a solution for the chosen wedge if
  $\operatorname{sgn}(\mathit{p}^{1}\mathit{p}^{0})=1$ and
  $|\mathit{p}^{1}|>|\mathit{p}^{0}|$.
  \par
  The above fixes the scalars on the horizon, but does not give us the
  $B$-coefficients; for that we need to solve eq.~(\ref{eq:29}), which for
  generic charges gives $B^{0}=s^{0}\mathit{p}^{0}$, where $s^{0}=\pm 1$ as is
  customary in this article. Given this identification we can then calculate
  the entropy density, whose positivity implies that
  $s^{0}=\operatorname{sgn}(\mathit{p}^{0})$:
  \begin{equation}
    \label{eq:60}
    \mathcal{S}^{3/2}(\mathit{p}) =
               |\mathit{p}^{0}\mathit{p}^{1}\mathit{p}^{2}|
               + \tfrac{\aleph^{2}}{3} |\mathit{p}^{0}|^{3} \, .
  \end{equation}
  The sign of $B^{0}$ together with the sign restrictions on the scalars
  determine the $B$'s to be $B^{I}=|\mathit{p}^{I}|$.
  \par
  Due to the fact that the magnetic charges must all have the same sign, the
  equations of motion (\ref{eq:8}) are satisfied for all values of $A^{I}$, so
  we can take $K^{I}=|A^{I}|+|\mathit{p}^{I}|\rho$ to ensure the regularity of
  the resulting solution and impose the normalization condition
  $\mathsf{V}(|A|)=1$ in order to obtain an asymptotic Minkowski metric. The
  final constraint comes from the fact that $\varphi^{1}(\rho )>1$: it is
  easily seen that this is satisfied if and only if $|A^{1}|>|A^{0}|$.
  \par
  A string in this class that saturates Gaida's bound, satisfies
  $\mathit{p}^{1}\mathit{p}^{2}=\aleph^{2}|\mathit{p}^{0}|^{2}$, and the
  resulting entropy density is
  \begin{equation}
    \label{eq:62}
    \left. \mathcal{S}\right|_{\text{Gaida}} = \left(\tfrac{4}{3}\right)^{2/3} |\mathit{p}^{0}|^{2} \, .
  \end{equation}
\item[b)] The second solution has no \textit{classical}, {\em
    i.e.\/}~$\aleph^{2}\rightarrow 0$, limit and exists if and only if
  $\operatorname{sgn}(\mathit{p}^{1}\mathit{p}^{2}) =1$. It reads
  \begin{equation}
    \label{eq:58}
    \varphi^{1}_{\rm h} = \frac{B^{1}}{B^{0}} = \sqrt{\frac{|\mathit{p}^{1}|}{|\mathit{p}^{2}|} \aleph^{2} }\, ,
    \qquad
    \varphi^{2}_{\rm h} = \frac{B^{2}}{B^{0}} = \sqrt{\frac{|\mathit{p}^{2}|}{|\mathit{p}^{1}|} \aleph^{2} } \, ,
  \end{equation}
  which lies in the desired wedge if $|\mathit{p}^{1}|> |\mathit{p}^{2}|$.
  Observe that this solution saturates Gaida's bound on the horizon.
  \par
  The Hamiltonian constraint on the horizon fixes
  \begin{equation}
    \label{eq:61}
    B^{0} = \tfrac{1}{2} s^{0}\operatorname{sgn}(\mathit{p}^{0}) \left(
            |\mathit{p}^{0}| + \operatorname{sgn}(\mathit{p}^{0}\mathit{p}^{1})
                \sqrt{ |\mathit{p}^{1}\mathit{p}^{2}| }
       \right) .
  \end{equation}
  The equations of motion show that they are satisfied iff
  $\mathit{p}^{1}\mathit{p}^{2}=\aleph^{2} |\mathit{p}^{0}|^{2}$, which
  immediately reduces this case to the Gaida solution of case a).
\end{itemize}
\par
The extremal non-BPS solutions to this case are not as easy to find as in the
$STU$ model,\footnote{ The system has a discrete symmetry with respect to the
  interchange of indices $1$ and $2$. The function $\mathcal{V}$ has the more
  important scaling symmetry $h^{1}\rightarrow e^{\lambda}h^{1}$ and
  $h^{2}\rightarrow e^{-\lambda}h^{2}$, but it does not leave the equations of
  motion in the H-formalism invariant.  } and one has to resort to a different
approach: first we solve the equations (\ref{eq:194}) in order to find the
relation between the $B$'s and the $\mathit{p}$'s and then solve the full
equations of motion. Clearly, solving the first of eqs.~(\ref{eq:194}) for the
$B$'s is challenging, but seeing that it is quadratic in $\mathit{p}$'s, we
first solve it to obtain $\mathit{p}=\mathit{p}(B)$ and then try to invert this
relation.
\par
Now there are four cases that solve the first of the eqs.~(\ref{eq:194}), one
of which corresponds to the BPS solution above and three correspond to extremal
non-BPS solutions:
\begin{itemize}
\item[i)] The first case is given by
  \begin{equation}
    \label{eq:64}
    \varphi^{1}_{\rm h} = \frac{B^{1}}{B^{0}} = -\frac{\mathit{p}^{1}}{\mathit{p}^{0}} 
           - \frac{2\aleph^{2}\mathit{p}^{0}}{3\mathit{p}^{2}} \, ,
    \qquad
    \varphi^{2}_{\rm h} = \frac{B^{2}}{B^{0}} = \frac{\mathit{p}^{2}}{\mathit{p}^{0}} \, .
  \end{equation}
  For this solution to be valid in the chosen wedge we must have that
  $\operatorname{sgn}(\mathit{p}^{0})=\operatorname{sgn}(\mathit{p}^{2})=-\operatorname{sgn}(p^{1})$
  and the magnetic charges must be such that
  \begin{equation}
    \label{eq:74}
    |\mathit{p}^{1}|>|\mathit{p}^{0}|+\frac{2|\mathit{p}^{0}|^{2}}{3|\mathit{p}^{2}|} \, .
  \end{equation}
  The normalization condition then gives $(\mathit{p}^{0})^{2}=(B^{0})^{2}$ and
  we can calculate
  \begin{equation}
    \label{eq:198}
    \mathcal{S}^{3/2} = B^{0} \left(
      |\mathit{p}^{1}\mathit{p}^{2}|-\tfrac{\aleph^{2}}{3}|\mathit{p}^{0}|^{2} \right) .
  \end{equation}
  As one can see, due to the restriction on the charges, we have that the term
  between the parentheses is positive, so we need to choose
  $B^{0}=|\mathit{p}^{0}|$, implying
  \begin{equation}
    \label{eq:65}
    \mathcal{S}^{3/2} = |\mathit{p}^{0}\mathit{p}^{1}\mathit{p}^{2}| - 
            \tfrac{\aleph^{2}}{3} |\mathit{p}^{0}|^{3} \, .
  \end{equation}
  Surprisingly, the Hamiltonian constraint does not impose any condition on
  $A^{1}$ but imposes the condition $\mathit{p}^{2}A^{0} = p^{0}A^{2}$. We
  solve this condition by introducing a positive number $\beta$ and writing
  \begin{equation}
    \label{eq:200}
    A^{0} = \beta |\mathit{p}^{0}| \, , \qquad 
    A^{2} = \beta |\mathit{p}^{2}| \, , \qquad\mbox{thus}\qquad
    \varphi^{2}(\rho ) = \frac{|\mathit{p}^{2}|}{|\mathit{p}^{0}|} \, .
  \end{equation}
  Even more surprisingly, the equations of motion are identically satisfied by
  the above relations between the coefficients $A$ and $B$.
  \par
  With the above information we can then calculate the metrical function
  \begin{equation}
    \label{eq:66}
    e^{-3U} = |\mathit{p}^{0}|\left(\beta+\rho\right)^{2} \left[
          |\mathit{p}^{2}|A^{1} + \tfrac{\aleph^{2}}{3}\beta|\mathit{p}^{0}|^{2}
        + \left( |\mathit{p}^{1}\mathit{p}^{2}|-\tfrac{\aleph^{2}}{3}|\mathit{p}^{0}|^{2} \right) \!\rho
    \right] .
  \end{equation}
  Its regularity becomes more manifest when we impose the asymptotic
  Minkowskianity condition
  \begin{equation}
    \label{eq:67}
    |\mathit{p}^{2}|A^{1} + \tfrac{\aleph^{2}}{3}\beta|\mathit{p}^{0}|^{2} = \frac{1}{\beta^{2}|\mathit{p}^{0}|} \, ,
  \end{equation}
  which allows us to express the metrical factor as
  \begin{equation}
    \label{eq:68}
    e^{-3U} = \left( 1 + \beta^{-1}\rho\right)^{2} \left[
                     1 + \beta^{2}
                     \left( |\mathit{p}^{0}\mathit{p}^{1}\mathit{p}^{2}|-\tfrac{\aleph^{2}}{3}|\mathit{p}^{0}|^{3} \right) \!\rho
              \right] .
  \end{equation}
  The tension of this string is easily calculated to give
  \begin{equation}
    \label{eq:69}
    4 \mathcal{T}_{(1)} = 2\beta^{-1} + \beta^{2} \left(
                  |\mathit{p}^{0}\mathit{p}^{1}\mathit{p}^{2}|-\tfrac{\aleph^{2}}{3}|\mathit{p}^{0}|^{3}
            \right) 
      = 2\beta^{-1} + \beta^{2} \mathcal{S}^{3/2}\, , 
  \end{equation}
  which is always positive, due to the restrictions imposed on the magnetic
  charges; the minimal attainable tension occurs when
  $\beta=\mathcal{S}^{-1/2}$, from which we have that $\mathcal{T}_{(1)} \geq
  \tfrac{3}{4}\mathcal{S}^{1/2}$.
  \par
  We have seen that $\varphi^{2}(\rho )$ is just a constant and that it always
  satisfies $\varphi^{2}>0$. The situation with $\varphi^{1}$ is slightly more
  complicated as it must satisfy $\varphi^{1}(\rho )>1$.  Writing out the
  constraint we see that
  \begin{equation}
    \label{eq:70}
    \beta|\mathit{p}^{0}|^{2}\left( |\mathit{p}^{2}| + \tfrac{\aleph^{2}}{3}|\mathit{p}^{0}|\right)
    - \beta^{-2} \leq
   \rho |\mathit{p}^{0}\mathit{p}^{2}| \left(
            |\mathit{p}^{1}|-|\mathit{p}^{0}|-\frac{2|\mathit{p}^{0}|^{2}}{3|\mathit{p}^{2}|}
       \right) .
  \end{equation}
  Since the term in brackets on the right-hand side is positive due to the
  condition for the scalar on the horizon to be in the correct wedge and since
  $\rho\in [0,\infty )$, we see that the left-hand side must in fact be smaller
  than zero, or
  \begin{equation}
    \label{eq:71}
    \beta^{3} \left( |\mathit{p}^{0}|^{2} |\mathit{p}^{2}| + \tfrac{\aleph^{2}}{3}|\mathit{p}^{0}|^{3}\right) < 1 \, .
  \end{equation}
\item[ii)] This case is readily obtained from case i) by using the obvious
  symmetry of the equations of motion and the Hamiltonian constraint under the
  interchange of the indices $1$ and $2$. What is not invariant under this
  change is the choice of wedge, which means that the restrictions we need to
  impose will be different from the ones imposed in case i). The solution to
  eq.~(\ref{eq:194}) is
  \begin{equation}
    \label{eq:72}
    \varphi^{1}_{\rm h} = \frac{B^{1}}{B^{0}} = \frac{\mathit{p}^{1}}{\mathit{p}^{0}} \, ,
    \qquad
    \varphi^{2}_{\rm h} = \frac{B^{2}}{B^{0}} = -\frac{\mathit{p}^{2}}{\mathit{p}^{0}} 
           - \frac{2\aleph^{2}\mathit{p}^{0}}{3\mathit{p}^{1}} \, .
  \end{equation}
  The choice of wedge then implies that
  $\operatorname{sgn}(\mathit{p}^{1})=\operatorname{sgn}(p^{0})$ and that
  $|\mathit{p}^{1}|>|\mathit{p}^{0}|$; the fact that $\varphi^{2}>0$ then
  implies that $\operatorname{sgn}(\mathit{p}^{2})=
  -\operatorname{sgn}(\mathit{p}^{0})$ and
  \begin{equation}
    \label{eq:73}
    |\mathit{p}^{1}| > |\mathit{p}^{0}|
    \qquad\mbox{and}\qquad
    |\mathit{p}^{2}| > \frac{2\aleph^{2}|\mathit{p}^{0}|^{2}}{3|\mathit{p}^{1}|} \, .
  \end{equation}
  The normalization condition in eq.~(\ref{eq:194}) gives
  $B^{0}=s^{0}\mathit{p}^{0}$ and as before the sign $s^{0}$ is fixed by the
  entropy density to be $s^{0}=\operatorname{sgn}(p^{0})$; the resulting
  entropy density is identical to the one in eq.~(\ref{eq:65}) and is positive
  owing to the restrictions (\ref{eq:73}).
  \par
  Similarly to what happened before, the Hamiltonian constraint and the
  equations of motion impose no condition on $A^{2}$, but impose the condition
  $A^{1}|\mathit{p}^{0}|=A^{0}|\mathit{p}^{1}|$. We solve it by
  $A^{0}=\gamma|\mathit{p}^{0}|$ and $A^{1}=\gamma|\mathit{p}^{1}|$, which
  immediately implies that $\varphi^{1}(\rho
  )=|\mathit{p}^{1}|/|\mathit{p}^{0}| >1$, so there is no possibility of the
  solution leaving the chosen wedge of moduli space.
  \par
  Regularity of the warp factor in ensured by the asymptotic Minkowskianity
  condition, which not only fixes
  \begin{equation}
    \label{eq:75}
    \gamma^{2}|\mathit{p}^{0}\mathit{p}^{1}| A^{2} = 1 - \tfrac{\aleph^{2}}{3}\gamma^{3} |\mathit{p}^{0}|^{3}
  \end{equation}
  and brings the metrical factor to the form in eq.~(\ref{eq:68}), but also
  means that the tension of the string in this case is the one in
  eq.~(\ref{eq:69}). The final ingredient then is the condition that
  $\varphi^{2}(\rho )$ be strictly positive.  This condition is easily
  calculated and gives
  \begin{equation}
    \label{eq:76}
    \tfrac{\aleph^{2}}{3}\gamma^{3} |\mathit{p}^{0}|^{3} < 1 \, .
  \end{equation}

\item[iii)] The third non-supersymmetric solution to eqs.~(\ref{eq:194}) can be
  found by imposing $\mathit{p}^{1}B^{2}=\mathit{p}^{2}B^{1}$, from which one
  finds that $\varphi^{1}_{\rm h}$ must satisfy the fourth-order equation
\begin{equation}
  \label{eq:77}
    2p_{0}p_{1}^{2} (\varphi^{1}_{\rm h})^{4} + 7p_{1}^{2}p_{2} (\varphi^{1}_{\rm h})^{3} -
    3p_{0}p_{1}p_{2} (\varphi^{1}_{\rm h})^{2} - 3p_{1}p_{2}^{2} \varphi^{1}_{\rm h} -
    3p_{0}p_{2}^{2} = 0 \, .
\end{equation}
The solutions are however too intricate to be of any real use and this case
will, therefore, not be treated.

\end{itemize}
%%%%%%%%%%%%%%%%%%%%%%%%%%%%%%%%%%%%%%%%%%%%%%%%%%%%%%%%%%%%%%%%%%%%%

\section{Conclusions}
\label{conclusions}

We have extended the H-FGK formalism of \cite{Meessen:2011aa} to black strings
and applied it to find examples of black-hole and black-string solutions in
specific models, as well as re-derive a non-extremal solution with constant
scalars discussed earlier by \cite{Mohaupt:2010fk}, which is a solution to any
model of $N=2$ supergravity in five dimensions, coupled to vector
multiplets. Since strings couple magnetically rather than electrically to the
gauge fields, the r\^{o}les of primary (untilded) and dual (tilded)
$H$-variables are interchanged in comparison with the case of black holes (for
distinction we denoted the primary variables $K$ when discussing strings). In
the $STU$ model, however, the resulting equations for black strings are the
same as for black holes.

The model-independent relationship between a set of parameters appearing in the
H-formalism and the asymptotic values of the scalars is a significant
simplification with respect to the original FGK formulation (in physical
variables), where the parameters need to be determined in each case from
complicated equations. For extremal solutions, the other set of parameters,
which can be called fake charges, is given by the condition that the black hole
potential be stationary, this way completing a simple procedure for
constructing extremal (supersymmetric and non-supersymmetric) black hole
solutions.

The derivation of first-order flow equations for non-supersymmetric extremal
black holes and black strings presented here allows the relation between the
fake and actual charges to be non-linear, which is indeed the case in the
specific example of the model from a Jordan sequence. For non-extremal
solutions, the hyperbolic ansatz makes it possible to bring the flow equations
to the same form as the extremal flow. On the other hand, one could argue that
once the harmonicity or hyperbolicity assumptions have been adopted, the
analysis of flow equations as such becomes perhaps superfluous, since the
radial profile of the scalars is already established by the respective
ans\"{a}tze.

We have demonstrated that for the $STU$ model in five dimensions the hyperbolic
(or exponential) ansatz in the non-extremal case and the harmonic ansatz in the
extremal case correspond to the most general solutions of the equations of
motion. We expect these ans\"{a}tze for the variables $H$ and $K$ to be valid
in all five- and four-dimensional models for all static solutions with
transverse spherical symmetry. For non-static black holes a less restrictive
ansatz is required, as the four-dimensional stabilization equations for general
extremal black holes suggest \cite{Galli:2010mg}.

%%%%%%%%%%%%%%%%%%%%%%%%%%%%%%%%%%%%%%%%%%%%%%%%%%%%%%%%%%%%%%%%%%%%%%
%%%%%%%%%%%%%%%%%%%%%%%%%%%%%%%%%%%%%%%%%%%%%%%%%%%%%%%%%%%%%%%%%%%%%%
%%%%%%%%%%%%%%%%%%%%%%%%%%%%%%%%%%%%%%%%%%%%%%%%%%%%%%%%%%%%%%%%%%%%%%
%%%%%%%%%%%%%%%%%%%%%%%%%%%%%%%%%%%%%%%%%%%%%%%%%%%%%%%%%%%%%%%%%%%%%%

\section*{Acknowledgments}

This work has been supported in part by the Spanish Ministry of Science and
Education grant FPA2009-07692, a Ram\'on y Cajal fellowship RYC-2009-05014, the
Princip\'au d'Asturies grant IB09-069, the Comunidad de Madrid grant HEPHACOS
S2009ESP-1473, and the Spanish Con\-solider-Ingenio 2010 program CPAN
CSD2007-00042. The work of C.S.S. has been supported by a JAE-predoc grant
JAEPre 2010 00613. T.O. wishes to thank M.M.~Fern\'andez for her unfaltering
support.

%%%%%%%%%%%%%%%%%%%%%%%%%%%%%%%%%%%%%%%%%%%%%%%%%%%%%%%%%%%%%%%%%%%%%%
%%%%%%%%%%%%%%%%%%%%%%%%%%%%%%%%%%%%%%%%%%%%%%%%%%%%%%%%%%%%%%%%%%%%%%
%%%%%%%%%%%%%%%%%%%%%%%%%%%%%%%%%%%%%%%%%%%%%%%%%%%%%%%%%%%%%%%%%%%%%%
%%%%%%%%%%%%%%%%%%%%%%%%%%%%%%%%%%%%%%%%%%%%%%%%%%%%%%%%%%%%%%%%%%%%%%
%%%%%%%%%%%%%%%%%%%%%%%%%%%%%%%%%%%%%%%%%%%%%%%%%%%%%%%%%%%%%%%%%%%%%%
%%%%%%%%%%%%%%%%%%%%%%%%%%%%%%%%%%%%%%%%%%%%%%%%%%%%%%%%%%%%%%%%%%%%%%
%%%%%%%%%%%%%%%%%%%%%%%%%%%%%%%%%%%%%%%%%%%%%%%%%%%%%%%%%%%%%%%%%%%%%%
%%%%%%%%%%%%%%%%%%%%%%%%%%%%%%%%%%%%%%%%%%%%%%%%%%%%%%%%%%%%%%%%%%%%%%
%%%%%%%%%%%%%%%%%%%%%%%%%%%%%%%%%%%%%%%%%%%%%%%%%%%%%%%%%%%%%%%%%%%%%%
%%%%%%%%%%%%%%%%%%%%%%%%%%%%%%%%%%%%%%%%%%%%%%%%%%%%%%%%%%%%%%%%%%%%%%

%%%%%%%%%%%%%%%%%%%%%%%%%%%%%%%%%%%%%%%%%%%%%%%%%%%%%%%%%%%%%%%%%%%%%%
%
%
\end{document}